\documentclass[structabstract]{aa}
\usepackage{graphicx}
\usepackage{natbib}
\usepackage{txfonts}
\usepackage{epstopdf}
\bibpunct{(}{)}{;}{a}{}{,}
\begin{document}

\newcommand{\bl}{\left\langle{}B_l\right\rangle}
\newcommand{\sigb}{\sigma_{\bl}}
\newcommand{\bmu}{\bl/\mu}
\newcommand{\icn}{I_{\rm C}}
\newcommand{\ilc}{I_{\rm L}}
\newcommand{\icnqs}{I_{\rm C,QS}}
\newcommand{\ilcqs}{I_{\rm L,QS}}
\newcommand{\sicnqs}{\sigma_{\icnqs}}
\newcommand{\silcqs}{\sigma_{\ilcqs}}
\newcommand{\mqsicn}{\left\langle{}\icnqs\right\rangle}
\newcommand{\mqsilc}{\left\langle{}\ilcqs\right\rangle}
\newcommand{\cicn}{C_{\icn}}
\newcommand{\cilc}{C_{\ilc}}
\newcommand{\bmut}{\left(\bmu\right)_{\rm TH}}
\newcommand{\g}{\:{\rm G}}
\newcommand{\bmua}{\bmut<\bmu\leq50\g}
\newcommand{\bmub}{50\g<\bmu\leq100\g}
\newcommand{\bmuc}{100\g<\bmu\leq180\g}
\newcommand{\bmud}{180\g<\bmu\leq280\g}
\newcommand{\bmug}{500\g<\bmu\leq640\g}
\newcommand{\bmuh}{640\g<\bmu\leq800\g}
\newcommand{\mua}{0.94<\mu\leq1.00}
\newcommand{\mug}{0.36<\mu\leq0.50}
\newcommand{\bmuf}{\frac{\bl}{\mu}}
\newcommand{\bmuqs}{\bmu\leq10\g}
\newcommand{\bmuz}{\bmu=0\g}
\newcommand{\bz}{\bl=0\g}
\newcommand{\umax}{\mu_{\rm max}}
\newcommand{\cicnmax}{C_{\icn,{\rm max}}}
\newcommand{\cilcmax}{C_{\ilc,\mu=1}}
\newcommand{\scicnmax}{\cicnmax/\left(\bmu\right)}
\newcommand{\scilcmax}{\cilcmax/\left(\bmu\right)}
\newcommand{\mbmu}{\left\langle\bmu\right\rangle}

\title{Intensity contrast of solar network and faculae}
\author{K.~L.~Yeo\inst{\ref{inst1},\ref{inst2}}\and S.~K.~Solanki\inst{\ref{inst1},\ref{inst3}}\and N.~A.~Krivova\inst{\ref{inst1}}}
\institute{
Max-Planck Institut f\"{u}r Sonnensystemforschung, Max-Planck-Stra\ss{}e 2, 37191 Katlenburg-Lindau, Germany \\
\email{yeo@mps.mpg.de}
\label{inst1}
\and
Technische Universit\"a{}t Braunschweig, Institut f\"{u}r Geophysik und Extraterrestrische Physik, Mendelssohnstra\ss{}e 3, 38106 Braunschweig, Germany
\label{inst2}
\and
School of Space Research, Kyung Hee University, Yongin, 446-701 Gyeonggi, Korea
\label{inst3}
}
\date{Received 1 November 2012 / Accepted 20 December 2012}
\abstract{
}{
This study aims at setting observational constraints on the continuum and line core intensity contrast of network and faculae, specifically, their relationship with magnetic field and disc position.
}{
Full-disc magnetograms and intensity images by the Helioseismic and Magnetic Imager (HMI) on-board the Solar Dynamics Observatory (SDO) were employed. Bright magnetic features, representing network and faculae, were identified and the relationship between their intensity contrast at continuum and line core with magnetogram signal and heliocentric angle examined. Care was taken to minimize the inclusion of the magnetic canopy and straylight from sunspots and pores as network and faculae.
}{
In line with earlier studies, network features, on a per unit magnetic flux basis, appeared brighter than facular features. Intensity contrasts in the continuum and line core differ considerably, most notably, they exhibit opposite centre-to-limb variations. We found this difference in behaviour to likely be due to the different mechanisms of the formation of the two spectral components. From a simple model based on bivariate polynomial fits to the measured contrasts we confirmed spectral line changes to be a significant driver of facular contribution to variation in solar irradiance. The discrepancy between the continuum contrast reported here and in the literature was shown to arise mainly from differences in spatial resolution and treatment of magnetic signals adjacent to sunspots and pores.
}{
HMI is a source of accurate contrasts and low-noise magnetograms covering the full solar disc. For irradiance studies it is important to consider not just the contribution from the continuum but also from the spectral lines. In order not to underestimate long-term variations in solar irradiance, irradiance models should take the greater contrast per unit magnetic flux associated with magnetic features with low magnetic flux into account.
}
\titlerunning{Intensity contrast of solar network and faculae}
\authorrunning{Yeo et al.}
\maketitle

\section{Introduction}
\label{introduction}

Photospheric magnetic activity is the dominant driver of variation in solar irradiance on rotational and cyclical timescales \citep{domingo09}. Magnetic flux in the photosphere is partly confined to discrete concentrations of kilogauss strengths, generally described in terms of flux tubes \citep{stenflo73,spruit83}. The brightness excess, or contrast relative to the Quiet Sun, of flux tubes is strongly modulated by their size and position on the solar disc \citep[see][for a review]{solanki93a}. Within these magnetic concentrations, pressure balance dictates an evacuation of the interior and consequent depression of the optical depth unity surface \citep{spruit76}. The horizontal extent influences the effect of radiative heating from the surroundings through the side walls on the temperature structure and contrast \citep{spruit81,grossmann94}. The position on the solar disc changes the viewing geometry, and therefore the degree to which the hot walls are visible and the apparent contrast \citep{steiner05}. Models describing the counteracting effects on solar irradiance of dark sunspots, and bright network and faculae, characterizing the latter by the magnetic filling factor (related to number density) and position have been successful in reproducing more than 90\% of observed variation over multiple solar cycles \citep{wenzler06,ball12}. Other factors, such as inclination, internal dynamics, phase of evolution and surrounding convective motions affect the brightness excess of a given flux tube, but become less important when considering the overall behaviour of an ensemble as is the case with such models \citep{fligge00}. The same is assumed of flux tube size, which enters these models only very indirectly, though it is known to have a significant effect on contrast.

Evidently, the robust reconstruction of solar irradiance variation from models based on photospheric magnetic activity is contingent, amongst other factors, on a firm understanding of the radiant behaviour of magnetic elements, in particular the variation with size and position on the solar disc. While the radiant behaviour of sunspots is relatively well known \citep{chapman94,mathew07} and sufficiently described by current models \citep{maltby86,collados94,unruh99}, the converse is true of network and faculae, and constitutes one of the main uncertainties in current solar irradiance reconstructions. This is primarily due to the difficulty in observing such small-scale features, the detailed structure of which are only starting to be resolved \citep{lites04,lagg10} with instruments such as the Swedish 1-m Solar Telescope, SST \citep{scharmer03} and the Imaging Magnetograph eXperiment, IMaX \citep{martinezpillet11} on-board SUNRISE \citep{solanki10,barthol11}. As such, the relationship between radiance and size cannot, as yet, be studied directly. It is however appropriate and more straightforward to consider instead the relationship between apparent intensity contrast and magnetogram signal. Apart from small-scale magnetic fields observed in the quiet Sun internetwork \citep{khomenko03,lites08,beck09}, magnetic concentrations carrying more than a minimum amount of flux exhibit similar field strengths regardless of size \citep{solanki93b,solanki99}. Flux tubes also tend towards vertical orientation due to magnetic buoyancy. As flux tubes exhibit a narrow range of magnetic field strengths and are largely vertical, on average the magnetogram signal at a given image pixel approximately scales with the proportion of the resolution element occupied by magnetic fields. Also, although the relationship between magnetogram signal and distribution of flux tube sizes is degenerate (a given magnetogram signal can, for example, correspond to either a single flux tube or a concentration of numerous smaller ones), flux tube size appears, on average, to be greater where the magnetogram signal is greater \citep{ortiz02}.

Relatively few studies examining network and faculae contrast variation with magnetogram signal and position on the solar disc have been reported in the literature. The majority of studies from the past two decades employed high-resolution ($<0.5\:{\rm arcsec}$) scans made with ground-based telescopes. For example, the efforts of \citet{topka92,topka97} and \citet{lawrence93} with the Swedish Vacuum Solar Telescope (SVST) and of \citet{berger07} with the SST. These studies suffer from variable seeing effects introduced by the Earth's atmosphere and poor representation of disc positions, a limitation imposed by the relatively narrow FOVs (field-of-view). \citet{ortiz02} and \citet{kobel11} repeated the work of \citet{topka92,topka97} and \citet{lawrence93} utilising observations from spaceborne instruments, and in so doing avoided seeing effects. \citet{ortiz02} employed full-disc continuum intensity images and longitudinal magnetograms from the Michelson Doppler Imager (MDI) on-board the Solar and Heliospheric Observatory (SoHO). While full-disc MDI observations presented a more complete coverage of disc positions, allowing the authors to derive an empirical relationship relating contrast to heliocentric angle and magnetogram signal, the spatial resolution is significantly poorer than in the SVST studies (4 arcsec versus $\gtrsim0.3$ arcsec). \citet{kobel11} examined the relationship between contrast and magnetogram signal near disc centre using spectropolarimetric scans from the Solar Optical Telescope (SOT) on-board Hinode \citep{kosugi07}. In this instance, the spatial resolution (0.3 arcsec) is comparable.

In this paper we discuss continuum and line core intensity contrast of network and facular elements from full-disc observations by the Helioseismic and Magnetic Imager (HMI) on-board the Solar Dynamics Observatory (SDO) spacecraft \citep{schou12}, and their relationship with heliocentric angle and magnetogram signal. The aim here is to derive stringent observational constraints on the relationship between intensity contrast, and position on the solar disc and magnetic field. This will be of utility to solar irradiance reconstructions, especially as HMI data will increasingly be used for this purpose.

This study partly echoes the similar studies discussed above, in particular that by \citet{ortiz02} utilising MDI observations. It presents a significant extension of the effort by \citet{ortiz02} in that we examined the entire solar disc not just in the continuum but also in the core of the HMI spectral line (Fe I 6173 \AA). This is of particular relevance to solar irradiance variation given the observation that spectral line changes appear to have a significant influence on such variations \citep{mitchell91,unruh99,preminger02}. Both here and in the study by \citet{ortiz02}, network and facular elements were distinguished from quiet Sun by the magnetogram signal, and sunspots and pores by the continuum intensity. HMI magnetograms are significantly less noisy than MDI magnetograms, allowing us to achieve a similar magnetogram signal threshold while averaging over a much shorter period (315 seconds versus 20 minutes). Network and facular features evolve at granular timescales \citep[$\sim10\:{\rm minutes,}$][]{berger96,wiegelmann12}. It is pertinent to keep the averaging period below this in order to avoid smearing and loss of signal. HMI also has a finer spatial resolution (1 arcsec compared to 4 arcsec), allowing weaker unresolved features to be detected at the same noise level than with MDI. The finer resolution however, also renders intensity fluctuations from small-scale phenomena such as granulation and filamentation more severe, which complicates the clear segmentation of sunspots and pores.

In Sect. \ref{data} we briefly present the HMI instrument, the observables considered and the data set. The data reduction process by which we identified and derived the intensity contrast of network and facular features is detailed in Sect. \ref{reduction}. Following that we describe the results of our analysis of these measured contrasts (Sect. \ref{results}). In Sect. \ref{discussion} we discuss our findings in the context of earlier studies and of their relevance to facular contribution to solar irradiance variation, before presenting our conclusions in Sect. \ref{conclusion}.

\section{Method}
\label{method}

\subsection{SDO/HMI data}
\label{data}

SDO/HMI \citep{schou12} is designed for the continuous, full-disc observation of velocity, magnetic field and intensity on the solar surface. The instrument comprises two $4096\times4096$ pixel CCD cameras observing the Sun at a spatial resolution of 1 arcsec (corresponding to two pixels).  By means of a tunable Lyot filter and two tunable Michelson interferometers, the instrument records 3.75-second cadence filtergrams at various polarizations and wavelengths across the Fe I absorption line at 6173 \AA{}. 45-second cadence Dopplergrams, longitudinal magnetograms and intensity (continuum, line depth and width) images are generated from the filtergram sequence. For this work we considered the longitudinal magnetic field, continuum intensity and line depth observables. HMI is full-Stokes capable, however, at time of study, only 720-second cadence Stokes IQUV parameters and Milne-Eddington inversions were available. As argued in Sect. \ref{introduction}, for this study it is important to keep the integration period of measurements below $\sim10\:{\rm minutes}$. We opted to utilise the 45-second longitudinal magnetograms also to keep in line with earlier studies, which examined intensity contrast variation with line-of-sight magnetic field \cite[e.g.][]{topka92,topka97,lawrence93,ortiz02,kobel11}. More details on the instrument can be found in \citet{schou12}.

The data set comprises simultaneous longitudinal magnetograms, continuum intensity and line depth images from 15 high activity days in the period May 2010 to July 2011. From each day, for each observable we took the average of the seven 45-second cadence images from a 315-second period, each rotated to the observation time of the middle image to co-register. Aside from signal-to-noise considerations, this averaging is to suppress variance from p-mode oscillations. The dates and times of the employed observations are listed in Table \ref{obs_date_time}.

\begin{table}[h]
\caption{Observation date and time of the data set.}
\label{obs_date_time}
\centering
\begin{tabular}{cc}
\hline\hline
Date & Time \\
(year.month.date) & (hour.minute.second) \\
\hline
2010.05.04 & 00:03:00 \\
2010.06.11 & 00:00:00 \\
2010.07.24 & 00:05:15 \\
2010.08.11 & 00:00:00 \\
2010.09.02 & 00:00:00 \\
2010.10.25 & 00:00:00 \\
2010.11.13 & 00:00:00 \\
2010.12.04 & 00:00:00 \\
2011.01.01 & 00:00:00 \\
2011.02.14 & 00:00:00 \\
2011.03.08 & 00:00:00 \\
2011.04.15 & 00:00:00 \\
2011.05.30 & 00:00:00 \\
2011.06.02 & 00:00:00 \\
2011.07.18 & 00:12:00 \\
\hline
\end{tabular}
\tablefoot{The time listed is the nominal time, in International Atomic Time (TAI), of the middle cadence in the sequence of seven considered from each day.}
\end{table}

Longitudinal magnetograms describe the line-of-sight component of the average magnetic flux density over each resolution element. To first order, for a given flux tube of intrinsic magnetic field strength $B$, the unsigned longitudinal magnetogram signal, $\bl$, is $\alpha{}B|\cos\gamma|$, scaled by $\alpha$, the magnetic filling factor and $\gamma$, the inclination of the magnetic field from the line-of-sight. As mentioned in the introduction (Sect. \ref{introduction}), flux tubes tend towards vertical orientation. In this study we examined the overall properties of an ensemble of magnetic elements. Under this condition it is reasonable to assume that on average $\gamma\approx\theta$, the heliocentric angle, allowing us to employ the magnetogram signal as a proxy for $\alpha{}B$ via the quantity $\bmu$, where $\mu=\cos\theta$. This quantity also represents a first-order approximation of the unsigned average magnetic flux density over each resolution element. At time of writing, HMI Dopplergrams and longitudinal magnetograms were generated from just the first Fourier component of the filtergram sequence resulting in a $\sqrt{2}$ factor increase in photon noise from the optimal level \citep{couvidat11,couvidat12}.

Each line depth image was subtracted from the corresponding continuum intensity image to yield the line core intensity image. Hereafter we will denote the continuum and line core intensity $\icn$ and $\ilc$ respectively. At time of writing, HMI data products are generated from the filtergram sequence assuming a Gaussian form to the Fe I 6173 \AA{} line and delta filter transmission profiles. The effect of these approximations on Doppler velocity and longitudinal magnetic field measurements are accounted for, but not completely for the intensity observables \citep{couvidat11,couvidat12}. The impact on this study is assumed minimal as we are only interested in contrast relative to the local mean quiet Sun level.

\subsection{Data reduction}
\label{reduction}
\subsubsection{Magnetogram noise level}

\begin{figure}[h]
\resizebox{\hsize}{!}{\includegraphics{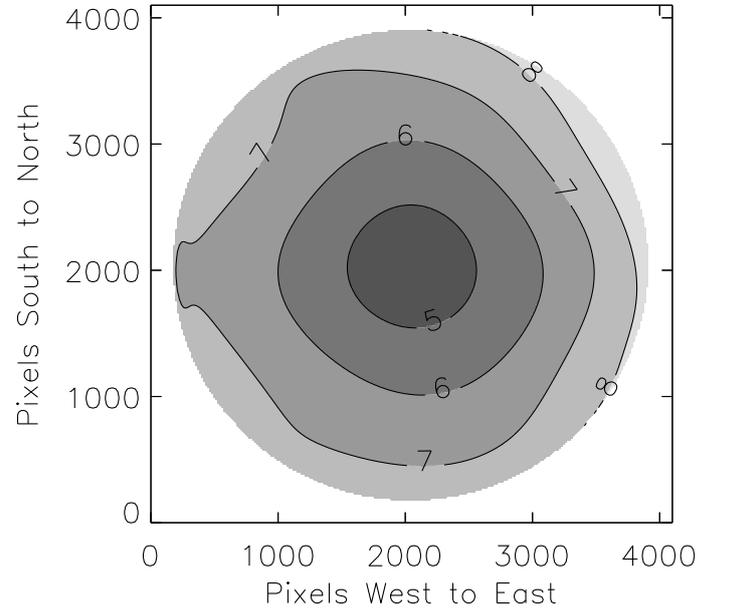}}
\caption{Magnetogram noise level in units of Gauss, as a function of disc position, sampled at 16-pixel intervals in either direction on the CCD.}
\label{noise_surface}
\end{figure}

The noise level of 315-second HMI magnetograms as a function of position on the solar disc was determined. For this purpose we used 10 spot-free 315-second magnetograms recorded over a seven month period in 2010. First, we estimated the centre-to-limb variation, CLV of the noise level. The pixels within each magnetogram were ordered by distance from the disc centre and sampled in successive blocks of 5000 pixels. The blocks represent concentric rings (near the limb, arcs, as the circumference of the solar disc is greater than 5000 pixels) of pixels of similar distance from the disc centre. Within each block we computed the average distance from the disc centre and the standard deviation of the magnetogram signal (iteratively, with points outside three standard deviations from the mean excluded from the succeeding iteration till convergence). A fifth-order polynomial in $\mu$ was fitted to the noise CLV; the mean of the standard deviation versus distance profiles so derived from the magnetograms. The magnetograms were then normalized by the noise CLV fit. At each disc position, the standard deviation over a $401\times401$ pixel window centred on the pixel of interest was computed (iteratively as above) for each normalized magnetogram and the median taken \citep[following][]{ortiz02}. (Near the limb, the standard deviation was computed from just the image pixels that lie within the solar disc.) A sixth-order polynomial was fitted to the resultant surface (termed here the noise residue). This fit represents the noise level after the removal of the CLV as a function of position on the solar disc. The noise level, $\sigb$, shown in Fig. \ref{noise_surface}, is then the product of the noise CLV fit and the noise residue fit. The noise level is lowest near disc centre and increases radially up to the limb (mean of $4.9\g$ for $\mu>0.95$ and $8.6\g$ for $\mu<0.05$). The root-mean-square, RMS difference between the noise level and the noise CLV fit is $0.4\g$. The correlation with distance from disc centre and relatively small deviation from circular symmetry suggests photon noise is the dominant component. The noise level of 45-second magnetograms, derived by a like analysis, has a similar, albeit accentuated form. The ratio between the noise level, averaged over the solar disc, of 45-second and 315-second magnetograms is 2.7 (approximately $\sqrt{7}$). The noise level of 45-second magnetograms determined by \citet{liu12}, via a vastly different method, also exhibits a similar CLV.

\subsubsection{Quiet Sun intensity}
\label{qsintensity}

For this part of the data reduction process, where we examined how quiet Sun intensity and the noise level of the intensity images vary with position on the solar disc, we counted all pixels with $\bmuqs$ as corresponding to quiet Sun (QS).

The continuum and line core intensity images were normalized by the fifth-order polynomial in $\mu$ fit to the quiet Sun pixels to correct for limb-darkening \citep[following][]{neckel94}. In the case of the line core intensity images, this was also to correct for the centre-to-limb broadening of the Fe I 6173 \AA{} line \citep{norton06}. There are distortions in the intensity images such that after this normalization, the mean quiet Sun intensity is not constant at unity but varying with position on the solar disc. The mean quiet Sun intensity at the continuum and line core as a function of position on the solar disc, denoted $\mqsicn$ and $\mqsilc$ respectively, were determined for each of the selected days. A $401\times401$ pixel window was centred on each disc position and the mean continuum and line core intensity of all quiet Sun pixels inside the window computed. For each day represented in the data set, $\mqsicn$ and $\mqsilc$ were given by the fifth-order polynomial fits to the mean quiet Sun continuum and line core intensity surfaces so derived from the images from the day. This analysis had to be repeated for each selected day as we found the spatial distribution of $\mqsicn$ and $\mqsilc$ to vary significantly over the period of observation. The RMS value of $\mqsicn-1$ and $\mqsilc-1$, the scale of the image distortions, is on average 0.004 and 0.01 respectively.

\begin{figure}[h]
\resizebox{\hsize}{!}{\includegraphics{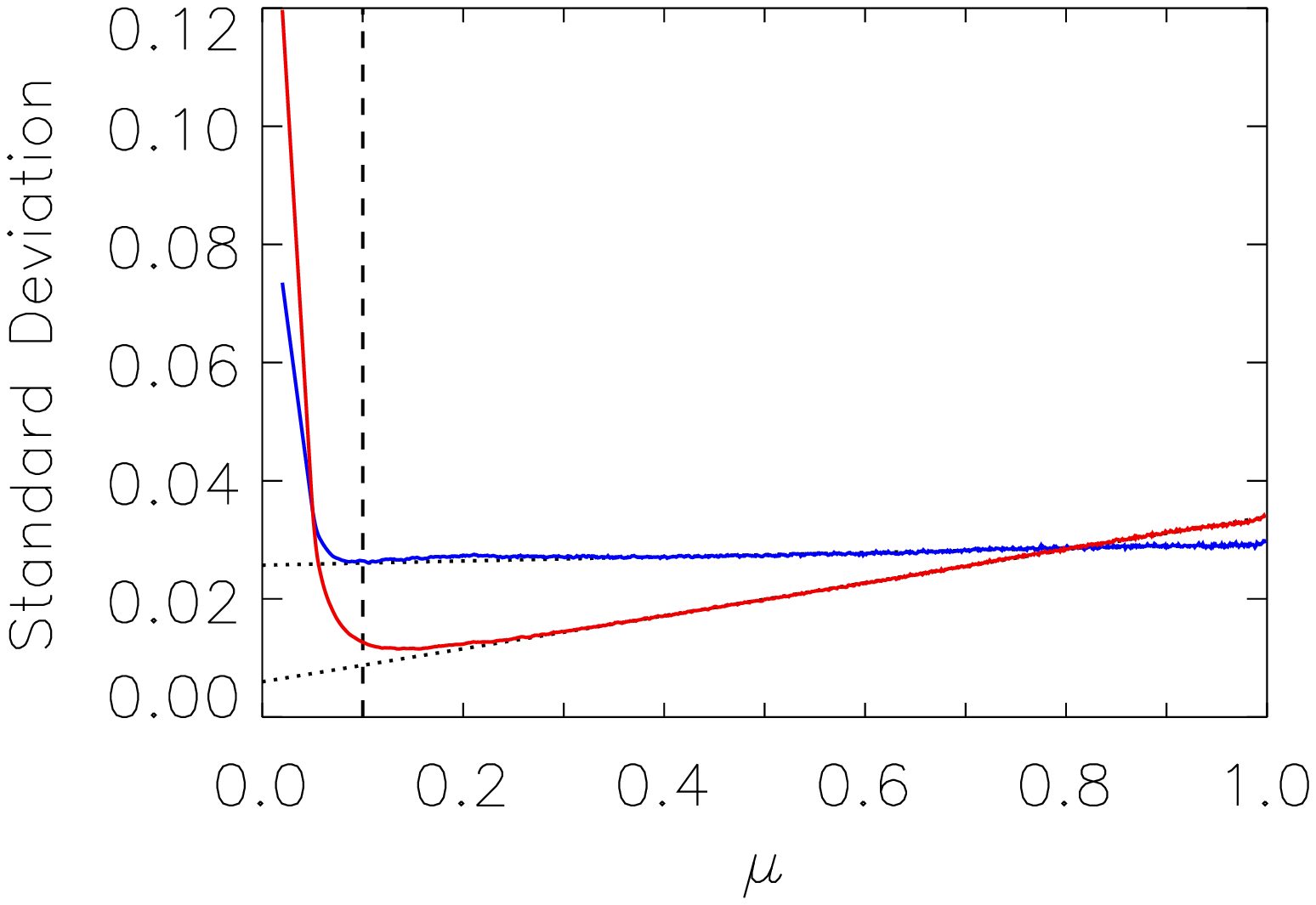}}
\caption{Standard deviation of quiet Sun ($\bmuqs$) intensity at continuum, $\sicnqs$ (red) and line core, $\silcqs$ (blue) as a function of $\mu$. The dotted lines represents the linear fit to $\sicnqs$ and $\silcqs$ over the range $0.3\leq\mu\leq1.0$ (largely hidden due to the close agreement), extrapolated to $\mu=0$. The dashed line denotes the threshold ($\mu=0.1$) below which pixels were excluded from the rest of the study in view of the scatter in measured intensity.}
\label{int_ld_noise_clv}
\end{figure}

The CLV of the standard deviation of quiet Sun intensity at continuum, $\sicnqs$ and line core, $\silcqs$ were determined, from $\icn/\mqsicn$ and $\ilc/\mqsilc$, by an analysis similar to the procedure used to derive the magnetogram noise CLV. In Fig. \ref{int_ld_noise_clv}, we express $\sicnqs$ and $\silcqs$, which carry information on the noise level of the intensity images and granulation contrast, as a function of $\mu$. Going from disc centre, $\sicnqs$ decreases gradually down to $\mu\sim0.15$ before increasing rapidly towards the limb. The monotonic decline from disc centre to $\mu\sim0.15$ resembles a similar trend in the CLV of granulation contrast reported by various authors \citep[][and references therein]{sanchezcuberes00,sanchezcuberes03}. $\silcqs$ exhibits a similar, though less accentuated, trend. For both $\sicnqs$ and $\silcqs$, the $\mu$-dependence from disc centre to $\mu\sim0.3$ is approximately linear, as highlighted by the linear fits (dotted lines). The elevation near limb is a direct consequence of limb-darkening; the diminishing signal-to-noise ratio translates into an escalating noise level in the normalized intensity. Given the gross scatter towards the limb, we excluded pixels outside $\mu=0.1$, representing about 1\% of the solar disc by area, from further consideration.

\subsubsection{Identification of network and faculae}

\begin{figure}[h]
\resizebox{\hsize}{!}{\includegraphics{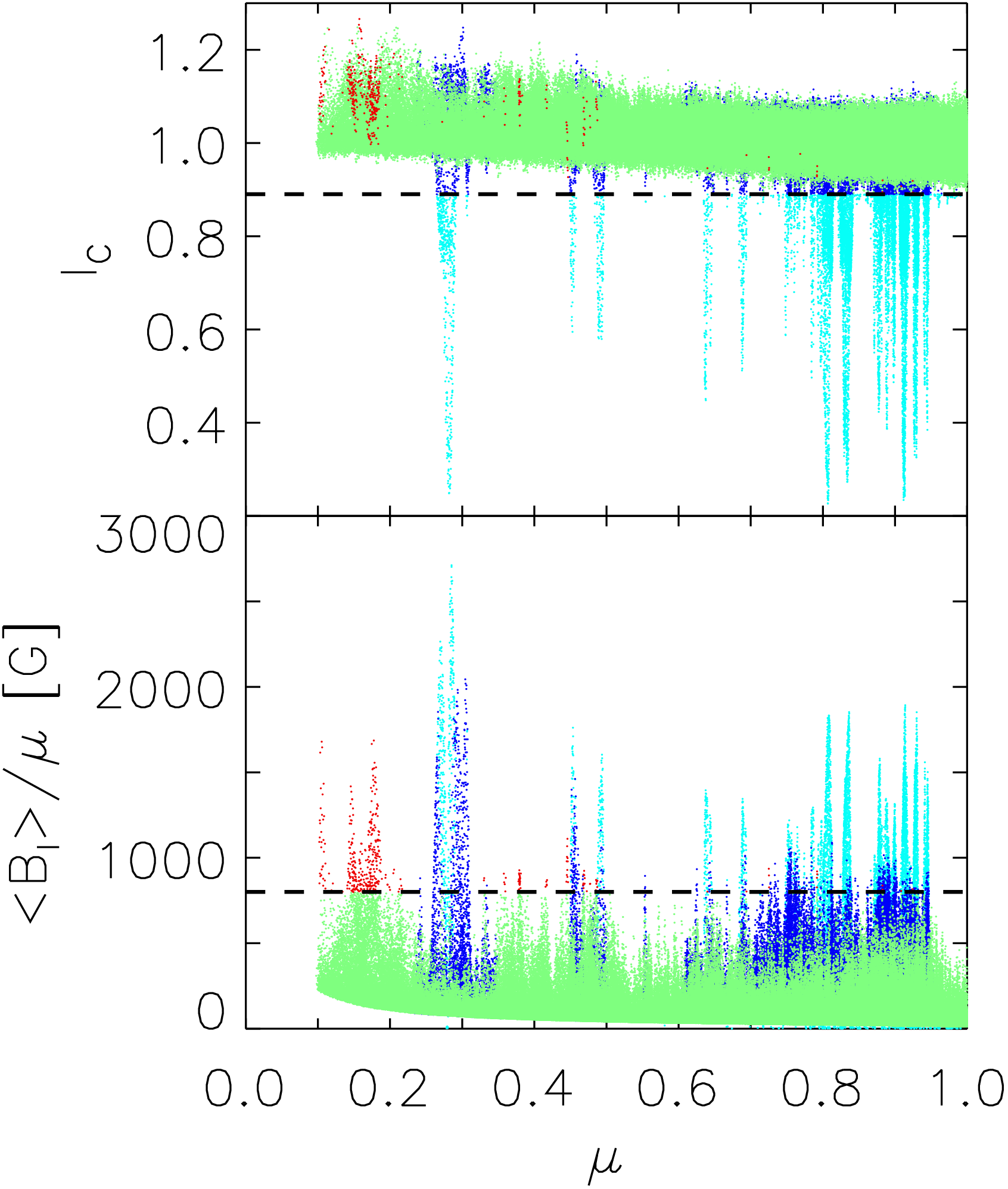}}
\caption{Continuum intensity, $\icn$ versus $\mu$ (top) and $\bmu$ versus $\mu$ (bottom) scatter plots of pixels counted as network and faculae (green) from June 2, 2011. The pixels counted as sunspots and pores by the continuum intensity threshold and the magnetic extension removal procedure (see text) are represented by the cyan and blue dots respectively. The red dots represent network and faculae pixels that lie above the cutoff $\bmu$ level. The dashed lines denote the continuum intensity threshold (top) and the cutoff $\bmu$ level (bottom).}
\label{fac_cnp_spt_int}
\end{figure}

Network and facular features, the subject of this work, were identified by first distinguishing them from quiet Sun by the magnetogram signal and from sunspots and pores by the continuum intensity. Pixels with $\bmu>3\sigb/\mu$ ($\sim14\g$ near disc centre, where $\sigb$ is lowest) were considered to harbour substantive magnetic fields. Isolated pixels above this threshold were assumed false positives and excluded. Hereafter we will denote $3\sigb/\mu$ as $\bmut$. Pixels with $\icn<0.89$ were counted as sunspots and pores. This value of the threshold is given by the mean of the minimum value of $\mqsicn-3\sicnqs$ from each selected day. It was so defined to distinguish sunspots and pores with minimal wrongful inclusion of intergranular lanes and magnetic features darker than the quiet Sun in the continuum. The continuum intensity versus $\mu$ scatter plot of network and facular pixels for one of the selected days (June 2, 2011) is shown in Fig. \ref{fac_cnp_spt_int} (top panel). The pixels identified as network and faculae lie clearly above the continuum intensity threshold.

\begin{figure}[h]
\resizebox{\hsize}{!}{\includegraphics{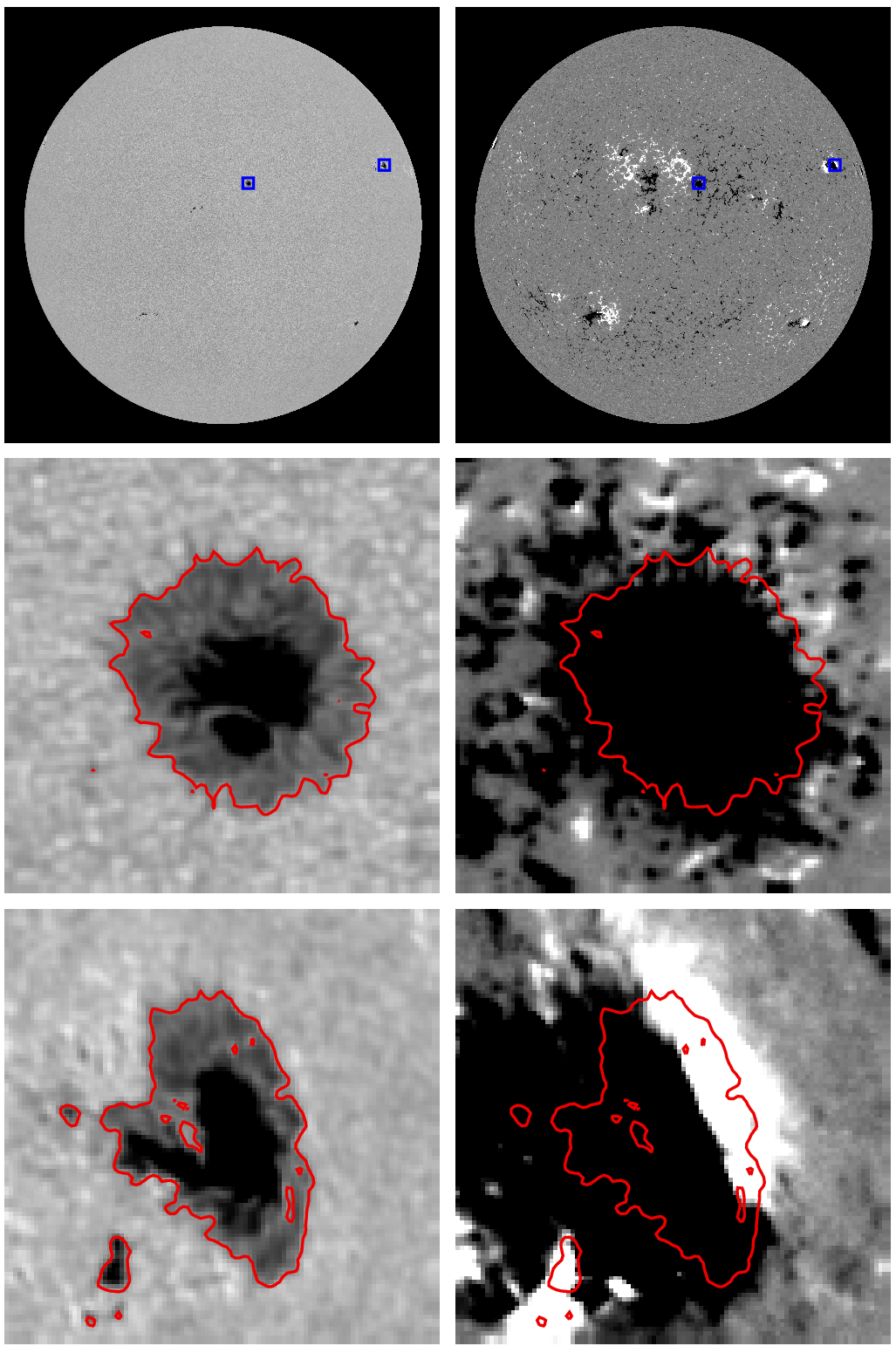}}
\resizebox{\hsize}{!}{\includegraphics{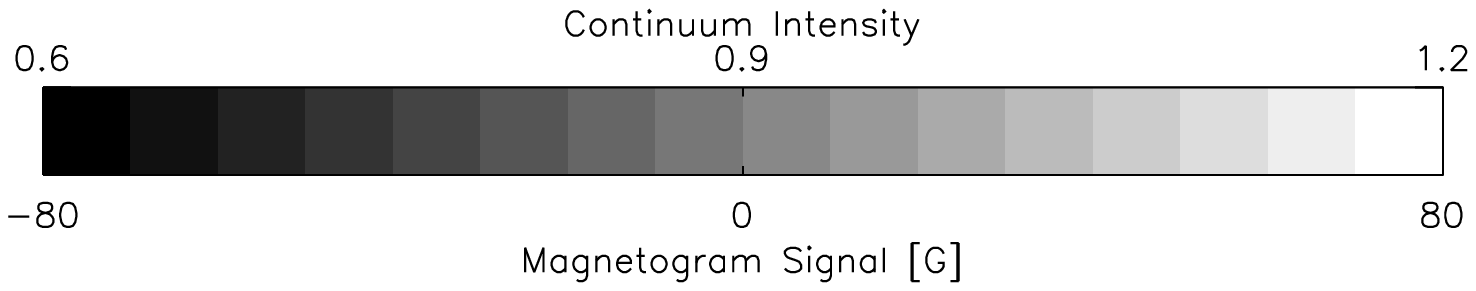}}
\caption{Continuum intensity image (top left) and magnetogram (top right) from July 18, 2011 and the corresponding $50\times50\:{\rm arcsec}$ insets of the boxed sunspot features near disc centre (middle panels) and limb (bottom panels). The red contours represent the continuum intensity sunspot boundary (the $\icn=0.89$ locus). The magnetograms have been saturated at $\pm80\g$ to highlight the extension of the magnetic signal from these sunspots beyond the intensity boundary (see grey scale below the figure).}
\label{canopy}
\end{figure}

In HMI magnetograms, magnetic signals produced by sunspots and pores extend beyond their boundary (in our analysis, the $\icn=0.89$ locus). This is illustrated for a sunspot near disc centre and another that is close to the limb on one of the selected days (July 18, 2011) in Fig. \ref{canopy}. The $\icn=0.89$ locus is plotted over the continuum intensity image and magnetogram of both sunspots (red contours) to highlight the extension of the magnetogram signal. This arises predominantly from the lateral expansion of their magnetic field with height (i.e., magnetic canopy) and partly also from the effect of straylight from sunspots and pores on nearby pixels. Towards the limb, these magnetic features become more extensive and bipolar due to the acute orientation of the largely horizontal magnetic canopy with the line-of-sight \citep{giovanelli80}, as illustrated by the near limb example (bottom panels). Figure \ref{canopy} also highlights the presence of bright sunspot structures (such as bright penumbral filaments) that lie above the continuum intensity threshold. All of these could easily be misidentified as network and faculae by the simple application of the magnetogram signal and continuum intensity thresholds described earlier.

To account for the effects of straylight around sunspots and pores, magnetic canopies and bright sunspot structures, we expanded sunspots and pores to include adjoining magnetic signal. Pixels adjacent to sunspots and pores that lie above $\bmut$ are reassigned to these features. This was iterated till no more pixels could be added. Here we will refer to this process as magnetic extension removal. Simply adding only adjoining pixels within a threshold distance from sunspots and pores instead is not useful as the physical extent of magnetic canopies exhibits a broad dynamic range, dependent on the position and physical properties of the associated feature. \citet{kobel11}, in a similar study with Hinode/SOT scans, expanded pores to include adjoining pixels above a threshold magnetogram signal level of $200\g$ to account for their influence on surrounding pixels from telescope diffraction. While this technique appears to work for Hinode/SOT scans, here we observed that adding only adjoining pixels above a threshold magnetogram signal level results in the appearance of a knee in contrast versus $\bmu$ plots at this threshold level regardless of the value chosen. For these reasons we expanded sunspots and pores to include all adjoining magnetic signal, though this conservative approach inevitably assigns too many pixels, including legitimate faculae, to sunspots and pores.

All pixels identified as magnetic, and not as sunspots and pores or their extensions, were taken to correspond to network and faculae. Figure \ref{fac_cnp_spt_int} shows the continuum intensity versus $\mu$ (top panel) and $\bmu$ versus $\mu$ (bottom panel) scatter plots of pixels identified as network and faculae (green), and counted as sunspots and pores by the magnetic extension removal procedure (blue) from June 2, 2011. The pixels captured by the magnetic extension removal procedure are not well distinguished from network and faculae by the continuum intensity; largely hidden by network and faculae in the continuum intensity versus $\mu$ scatter plot. It is clear from the $\bmu$ versus $\mu$ scatter plots however that the two classes are significantly different magnetic populations. As noted earlier, this procedure is likely too severe and results in the exclusion of some true faculae. However, for the purpose of this study it is not necessary to identify all faculae present and far more important to avoid false positives.

Finally, network and facular pixels with $\bmu$ above a conservative cutoff level of $800\g$ were excluded from the subsequent analysis \citep[following][]{ball11,ball12}. They are mostly bright features concentrated near the limb (as illustrated for June 2, 2011 by the red dots in Fig. \ref{fac_cnp_spt_int}) associated with sunspots and pores. (The relatively high $\bmu$ values likely reflect nearly horizontal fields, for which $|\cos\gamma|\gg\mu$ towards the limb.) This is to account for non-facular magnetic signals that might have been missed by the continuum intensity threshold and the magnetic extension removal procedure.

\begin{figure*}[t]
\centering
\includegraphics[width=17cm]{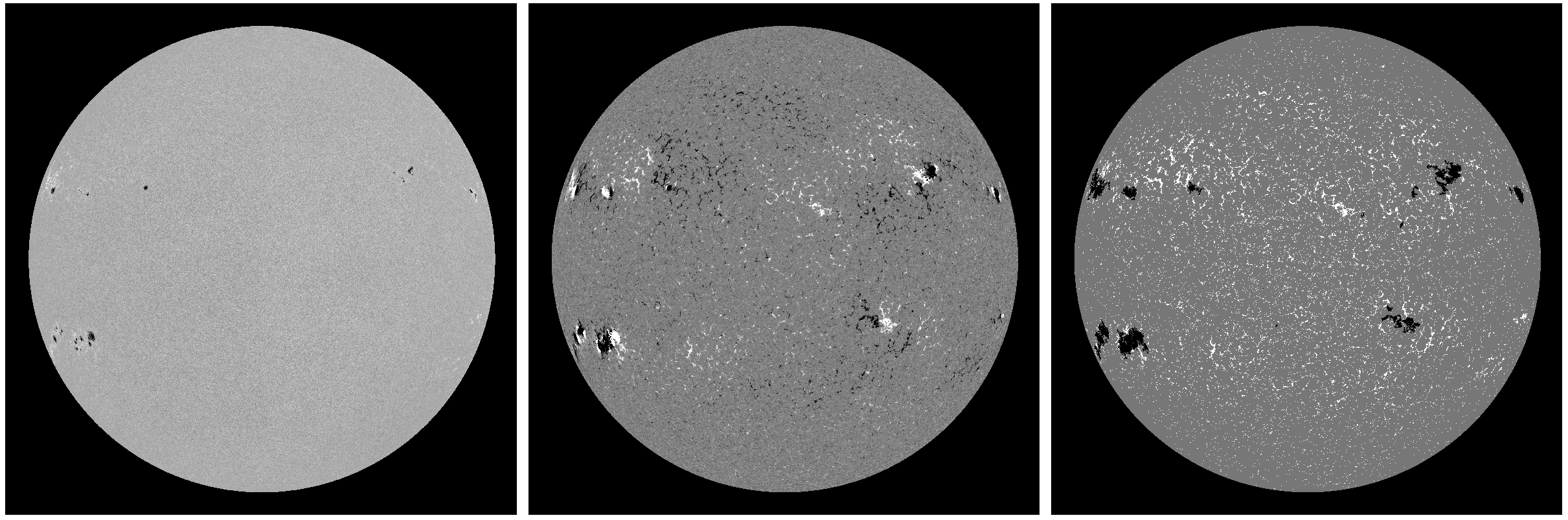}
\caption{HMI continuum intensity image (left), magnetogram (middle) and classification image (right) from May 30, 2011. The classification image indicates the positions of the pixels classed as network and faculae (white), and sunspots and pores (black). The latter includes magnetic signal adjoined to sunspots and pores, counted to them to avoid counting their magnetic canopy, possible bright structures within penumbrae and straylight as network and faculae erroneously. The continuum intensity image and magnetogram are scaled between 0.6 and 1.2, and $-80\g$ and $80\g$ respectively as in Fig. \ref{canopy}}
\label{network_plage_mask}
\end{figure*}

The classification image, indicating the positions of the pixels classed as sunspot and pores, and network and faculae for another of the selected days (May 30, 2011) is shown in Fig. \ref{network_plage_mask}. In spite of the severe measures taken to minimise the influence of sunspots and pores, a fair fraction of active region faculae remains. In total, $7.6\times10^6$ pixels were identified as corresponding to network and facular features from the data set (i.e. $4.5\%$ of all solar disc pixels in the 15 continuum intensity image and magnetogram pairs examined).

\subsubsection{Definition of intensity contrast}

The average continuum and line core intensity contrast over a given resolution element, $\cicn$ and $\cilc$ were defined as:
\begin{equation}
\cicn\left(x,y\right)=\frac{\icn\left(x,y\right)-\mqsicn(x,y)}{\mqsicn\left(x,y\right)}
\label{intensity_contrast}
\end{equation}
and
\begin{equation}
\cilc\left(x,y\right)=\frac{\ilc\left(x,y\right)-\mqsilc(x,y)}{\mqsilc\left(x,y\right)}
\label{linecore_contrast}
\end{equation}
respectively, where $\left(x,y\right)$ denote position on the CCD array. These two quantities, computed for each of the pixels identified as corresponding to network and facular features, represent the normalized difference between the continuum and line core intensities at a given pixel and the local mean quiet Sun levels as given by the mean quiet Sun continuum and line core intensity surfaces, $\mqsicn$ and $\mqsilc$, defined earlier in Sect. \ref{qsintensity}.

In summary, here we extracted an ensemble of $7.6\times10^6$ continuum and line core intensity contrast measurements corresponding to network and facular features covering as wide a range of heliocentric angles ($0.1\leq\mu\leq1.0$) and magnetogram signal ($\bmut<\bmu\leq800\g$) as reasonably possible from the data set for the succeeding analysis.

\section{Results}
\label{results}

\subsection{Variation with position and magnetogram signal}
\label{profile}

\begin{figure*}[h]
\centering
\includegraphics[width=17cm]{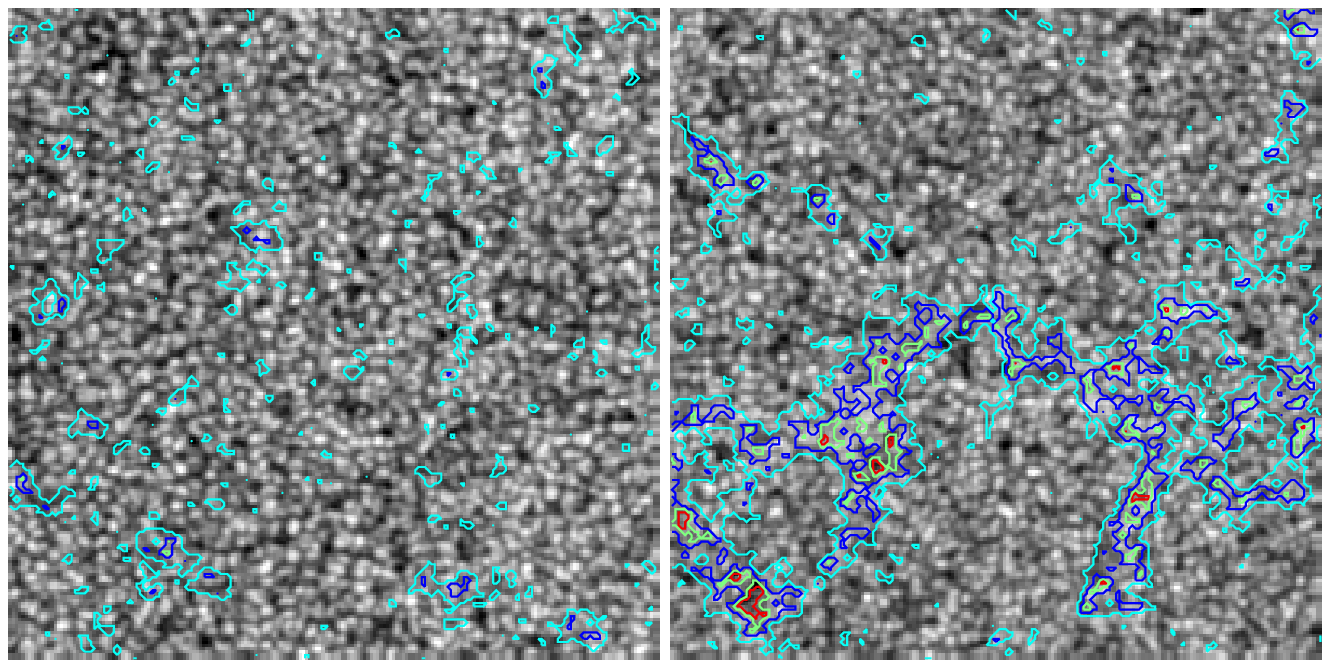}
\caption{$100\times100$ arcsec insets of a quiet region near disc centre ($\mu>0.99$, left) and active region NOAA 11187 ($0.82<\mu<0.91$, right) of the continuum intensity image from April 15, 2011. The cyan contours indicate the boundary of network and faculae. The blue, green and red contours correspond to $\bmu=100\g$, $280\g$ and $500\g$, respectively.}
\label{qs_ar_fac_map}
\end{figure*}

The positions of pixels identified as network and faculae, classified by $\bmu$, in a quiet region and an active region on one of the selected days (April 15, 2011) is shown in Fig \ref{qs_ar_fac_map}. As expected, at HMI's spatial resolution, magnetic signals with higher $\bmu$ are largely concentrated in active regions. Though magnetic signals with lower $\bmu$ are present in both quiet Sun and active regions, the fact that the solar disc is predominantly quiet Sun means these signals correspond largely to quiet Sun network and internetwork.

\begin{figure*}[h]
\centering
\includegraphics[width=17cm]{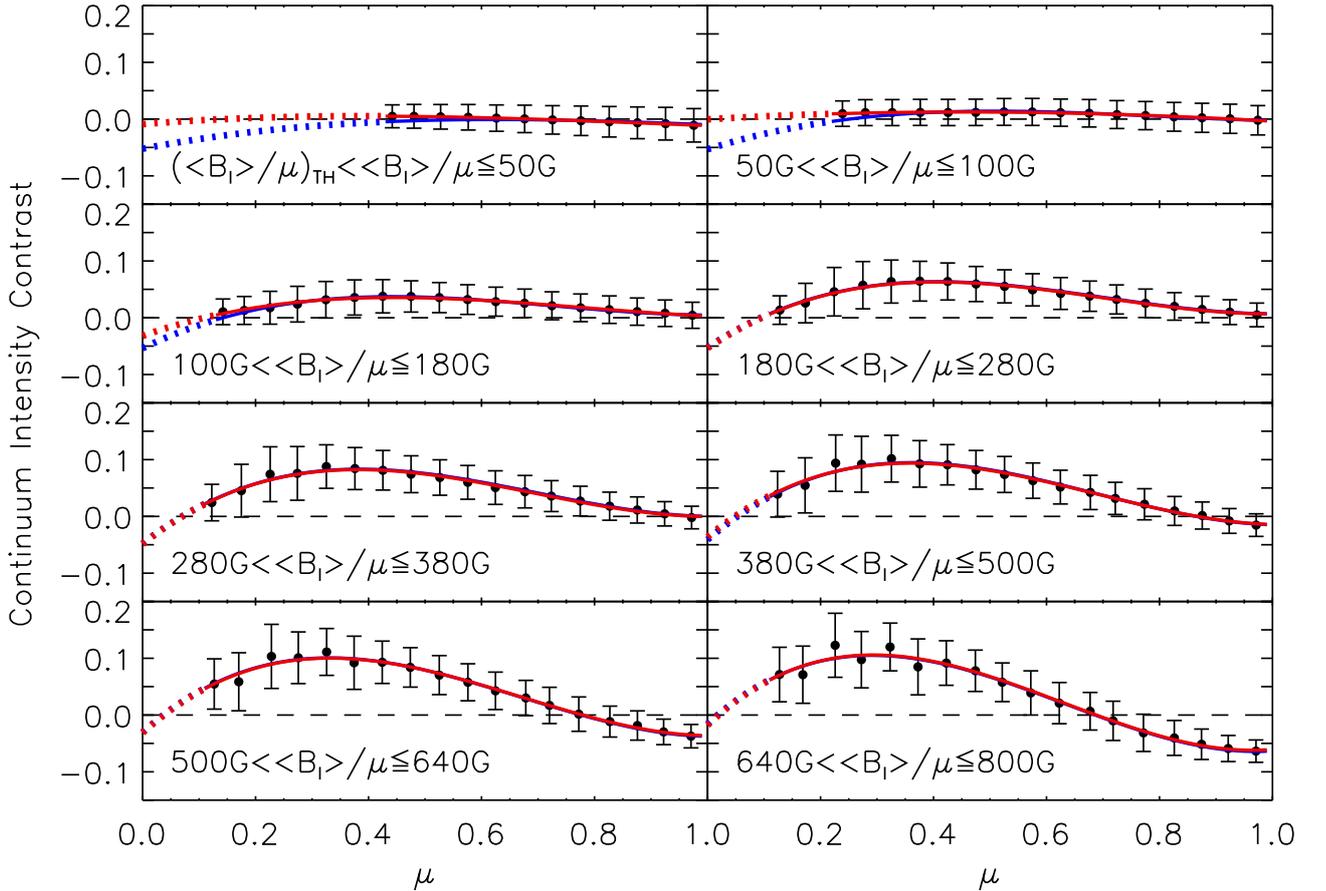}
\caption{CLV of continuum intensity contrast, $\cicn$ over eight $\bmu$ intervals. The filled circles and error bars represent the mean and standard deviation of $\cicn$ grouped by $\mu$ in bins 0.05 wide. The red curves are third-order polynomial fits to the filled circles and the blue curves are the cross-sections of the surface fit to $\cicn$ at the mean $\bmu$ within each interval (largely hidden due to the close agreement), extrapolated to $\mu=0$ (dotted segments). The horizontal dashed lines denote the mean quiet Sun level.}
\label{intensity_contrast_scatter_bmu}
\end{figure*}

\begin{figure*}[h]
\centering
\includegraphics[width=17cm]{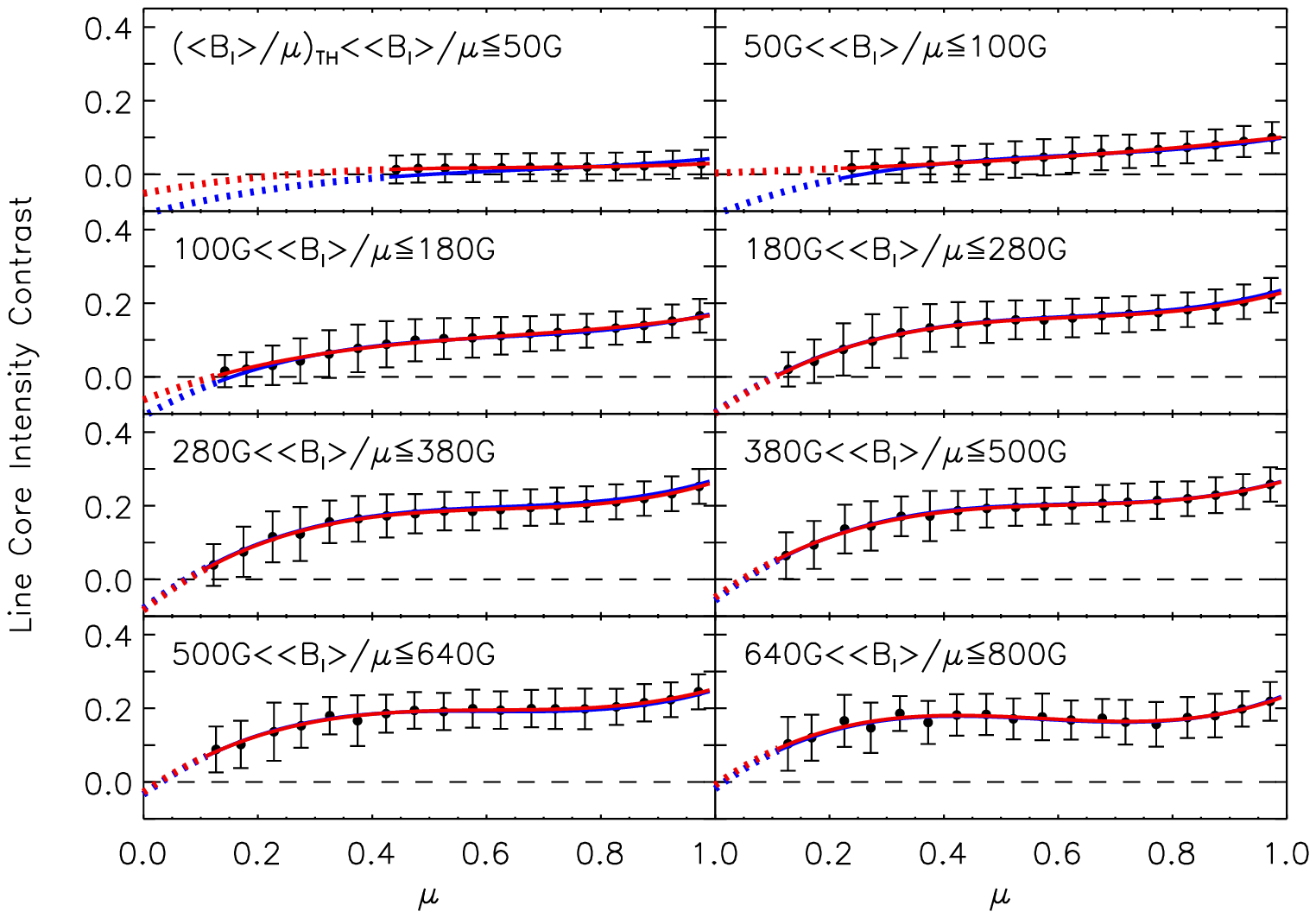}
\caption{Similar to Fig. \ref{intensity_contrast_scatter_bmu}, but for line core intensity contrast.}
\label{linecore_contrast_scatter_bmu}
\end{figure*}

To elucidate the CLV of intensity contrast, we grouped the measured contrasts into eight intervals of $\bmu$ spanning the range $\bmut<\bmu\leq800\g$ and within each interval into $\mu$ bins 0.05 wide. As the distribution of magnetogram signal is skewed towards the lower bound \citep{wenzler04,parnell09}, the $\bmu$ intervals were defined such that the widths slide from about $36\g$ ($\bmua$) to $160\g$ ($\bmuh$) to ensure reasonable statistics in every interval. In grouping the measured contrasts into these broad $\bmu$ intervals we are effectively grouping the network and facular features by $\alpha{}B$ (Sect. \ref{data}), neglecting differences in quiet Sun network and active region faculae contrast \citep{lawrence93,kobel11}. This is reasonable since the lower intervals are mainly populated by quiet Sun network and the higher intervals by active region faculae. The bin-averaged continuum and line core intensity contrast as a function of $\mu$ and the cubic polynomial fit for each of the $\bmu$ intervals are expressed in Figs. \ref{intensity_contrast_scatter_bmu} and \ref{linecore_contrast_scatter_bmu} respectively. For brevity we will refer to these bin-averaged contrasts as the contrast CLV profiles.

The truncated $\mu$ coverage in the lower $\bmu$ intervals is due to foreshortening. As flux tubes are mainly vertical, going from disc centre to the limb the corresponding longitudinal magnetogram signal diminishes, eventually dropping below the threshold level which itself rises towards the limb (Fig. \ref{noise_surface}). The fluctuations near the limb, more pronounced in the higher $\bmu$ intervals, is due to the inhomogeneous distribution of active regions on the solar disc, where stronger magnetic signals are concentrated on the selected days. Diminishing statistics also play a role; there are comparatively fewer pixels in the higher $\bmu$ intervals.

\begin{figure*}[h]
\centering
\includegraphics[width=17cm]{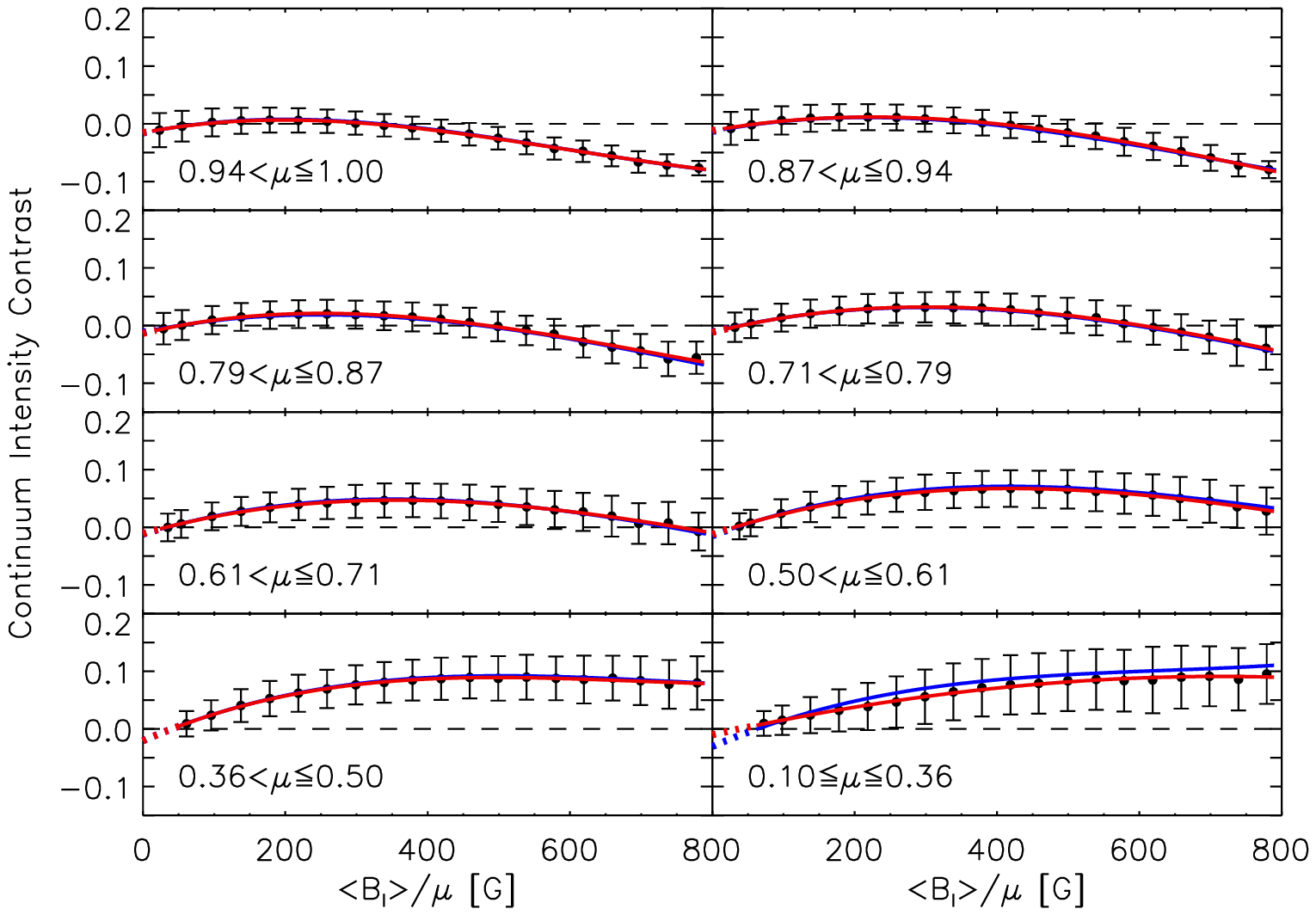}
\caption{Continuum intensity contrast, $\cicn$ as a function of $\bmu$ over eight $\mu$ intervals. The filled circles and error bars represent the mean and standard deviation of $\cicn$ grouped by $\bmu$ in $40\g$ bins. The red and blue curves and the horizontal dashed lines have the same meanings as in Fig. \ref{intensity_contrast_scatter_bmu}.}
\label{intensity_contrast_scatter_mu}
\end{figure*}

\begin{figure*}[h]
\centering
\includegraphics[width=17cm]{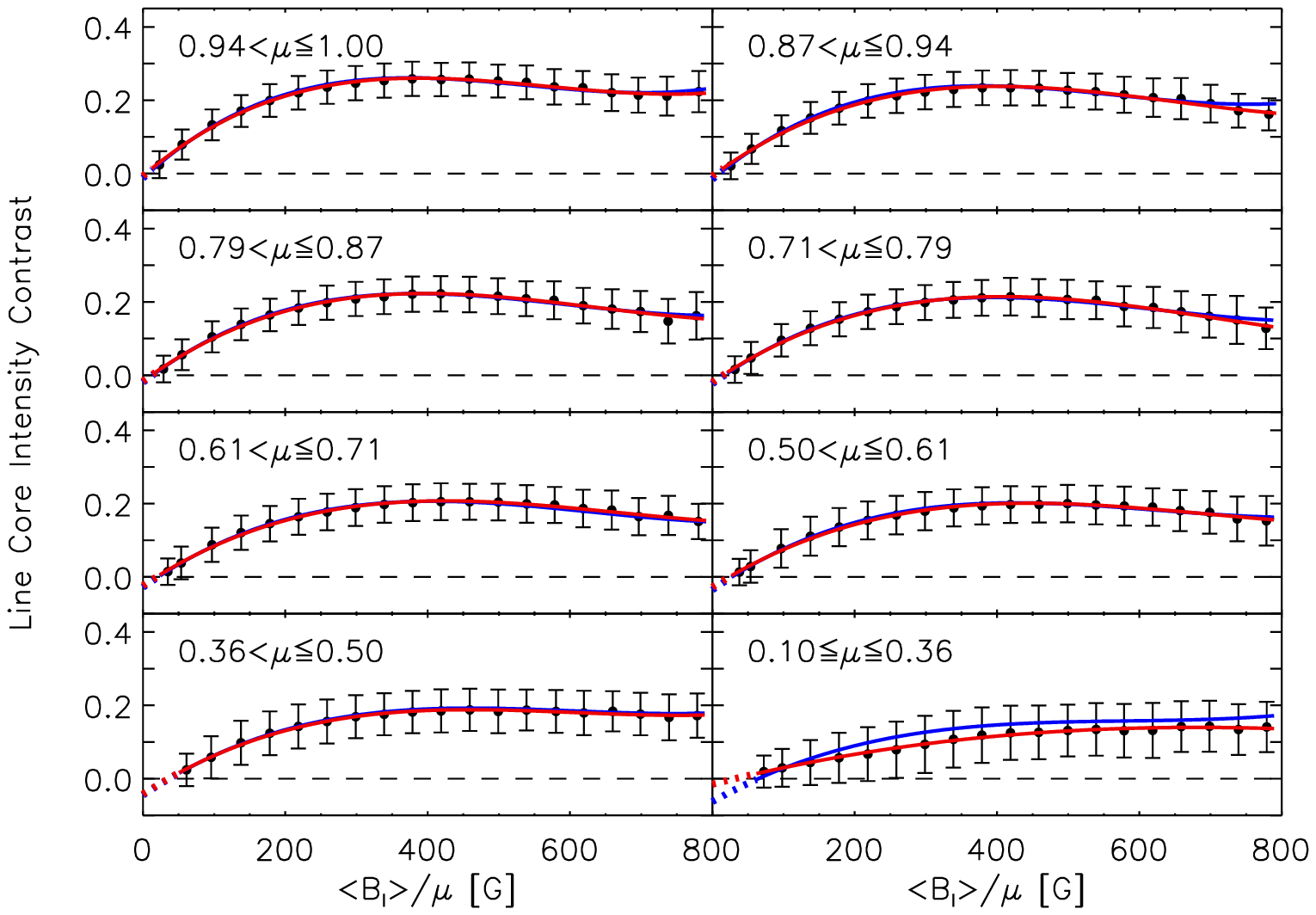}
\caption{Similar to Fig. \ref{intensity_contrast_scatter_mu}, but for line core intensity contrast.}
\label{linecore_contrast_scatter_mu}
\end{figure*}

To investigate the $\bmu$-dependence of intensity contrast, the measured contrasts were grouped into eight $\mu$ intervals spanning the range $0.1\leq\mu\leq1.0$ and within each interval into $\bmu$ bins $40\g$ wide. The $\mu$ intervals were defined such that they represent an approximately equal proportion of the solar disc by area. The bin-averaged continuum and line core intensity contrasts as a function of $\bmu$ and the cubic polynomial fit for each of the $\mu$ intervals are shown in Figs. \ref{intensity_contrast_scatter_mu} and \ref{linecore_contrast_scatter_mu} respectively. For brevity we will refer to these bin-averaged contrasts as the contrast versus $\bmu$ profiles. These profiles represent the variation of continuum and line core intensity contrast ranging from internetwork and weak network to active region faculae at different distances from disc centre.

The cubic polynomial fit to each contrast versus $\bmu$ profile included the zeroth-degree term. This produced better fits to data at low $\bmu$ than constraining the fits to pass through the origin by excluding it. Approaching $\bmuz$, the contrast versus $\bmu$ profiles decline to weak, broadly negative levels. Although the cubic polynomial fits express good agreement with measurements, they are too simple to accommodate this decline well without including the unphysical zeroth-degree term. For similar considerations we included the zeroth-degree term in the cubic polynomial fits to the contrast CLV profiles, so rendering them non-zero at $\mu=0$. These apparent offsets in the contrast CLV and contrast versus $\bmu$ profiles probably reflects the fact that magnetic elements are generally located in dark intergranular lanes.

\begin{figure}
\resizebox{\hsize}{!}{\includegraphics{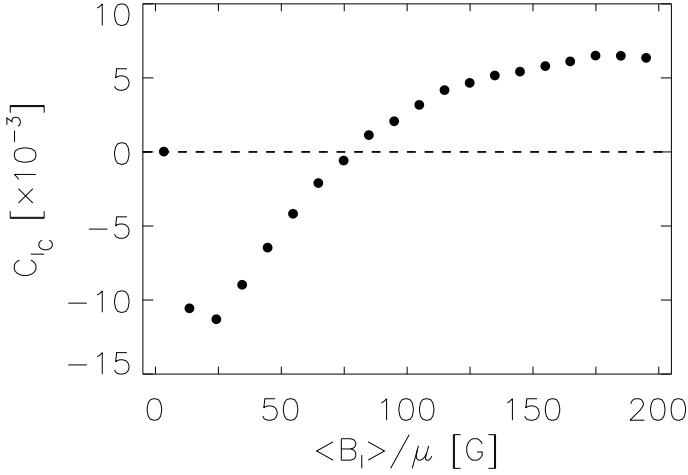}}
\caption{Continuum intensity contrast, $\cicn$ of quiet Sun, network and faculae over $\mua$ as a function of $\bmu$. The filled circles represent the mean of $\cicn$ grouped by $\bmu$ in bins $10\g$ wide. The dashed line denotes the mean quiet Sun level.}
\label{fishhook}
\end{figure}

In Fig. \ref{fishhook} we show a recomputation of the continuum intensity contrast versus $\bmu$ profile over $\mua$ from Fig. \ref{intensity_contrast_scatter_mu} where we included pixels below the magnetogram signal threshold and not identified as sunspots and pores (i.e., quiet Sun), and grouped the measurements by $\bmu$ in bins $10\g$ (instead of $40\g$) wide. Approaching $\bmuz$, contrast declines gradually to below the reference level before turning back up sharply towards the origin. Similar trends were reported by \citet{narayan10}, \citet{kobel11} and \citet{schnerr11}, who termed it the fishhook feature, based on SST and Hinode/SOT disc centre scans. \citet{schnerr11} demonstrated the resolution of granules and dark intergranular lanes, where magnetic flux concentrates, at the relatively fine spatial resolution of both instruments (0.15 and 0.3 arcsec respectively) to be the cause of the fishhook feature at low magnetogram signal levels. Though HMI has a coarser resolution (1 arcsec) than either Hinode/SOT or SST, the fishhook feature near $\bmuz$ in Fig. \ref{fishhook} indicates granulation is still sufficiently resolved to have a measurable impact on apparent contrast.

Comparing the measured intensity contrast at continuum and line core derived here, the most distinct difference is the opposite CLV. Confining this discussion to broad trends in the measurements, continuum intensity contrast is weakest near disc centre and increases up to a maximum before declining quite significantly towards the limb (Fig. \ref{intensity_contrast_scatter_bmu}). Conversely, line core intensity contrast is strongest near disc centre and declines monotonically from disc centre to limb (Fig. \ref{linecore_contrast_scatter_bmu}). Line core intensity is modulated by line strength and shape, and continuum intensity. The centre-to-limb decline exhibited by the measured line core intensity contrast arises from variation in line strength and shape (if this were absent the continuum and line core intensity contrast would vary proportionally) and would be even more acute than reflected in the CLV profiles if not partially offset by the accompanying increase in continuum intensity. This will be demonstrated, along with a closer discussion of the diverging trends exhibited by both sets of measurements in Sect \ref{compare_lc}.

\subsection{Surface fits}
\label{surface}

The cubic polynomial fit to the contrast CLV and contrast versus $\bmu$ profiles reproduced the observations well (Figs. \ref{intensity_contrast_scatter_bmu} to \ref{linecore_contrast_scatter_mu}). Given this, we fit the entire set of measured network and faculae continuum and line core intensity contrast as functions of $\mu$ and $\bmu$ \citep[following][]{ortiz02}. The measured contrasts were grouped by $\mu$ and the natural logarithm of $\bmu$ into a grid of $36\times41$ bins. The grid spans $0.1\leq\mu\leq1.0$ and $2.6\leq\ln\left(\bmu\right)\leq6.7$ or about $14\g$ to $810\g$. Each bin represents an interval of 0.025 in $\mu$ and 0.1 in $\ln\left(\bmu\right)$. The logarithmic binning in $\bmu$ is to compensate for the bottom-heavy distribution of magnetogram signal, to ensure the distribution of points is not too concentrated in the lower magnetogram signal bins. At each grid element we considered the mean $\mu$, $\bmu$, continuum and line core intensity contrast of the points within the bin. In accord with the cubic polynomial fits to the individual contrast CLV and contrast versus $\bmu$ profiles, bivariate polynomials cubic in $\mu$ and $\bmu$ were fitted to the bin-averaged continuum and line core intensity contrast. The zeroth $\mu$ and $\bmu$ orders were included based on similar considerations as with the polynomial fits to the contrast CLV and contrast versus $\bmu$ profiles. In linear algebra notation, the surface fit to the bin-averaged continuum and line core intensity contrast are:
\begin{eqnarray}
\lefteqn{\cicn\left(\mu,\bmuf\right)=}\nonumber\\
&&\left[\begin{array}{c}
10^{-2}\left(\bmuf\right)^0 \\ 10^{-3}\left(\bmuf\right)^1 \\ 10^{-6}\left(\bmuf\right)^2 \\ 10^{-9}\left(\bmuf\right)^3
\end{array}\right]^T
\left[\begin{array}{rrrr}
-5.11 & 7.74 & 0.34 & -4.72 \\ -0.04 & 3.84 & -7.42 & 3.90 \\ 0.19 & -6.27 & 12.03 & -6.78 \\ -0.08 & 3.58 & -8.04 & 5.04
\end{array}\right]
\left[\begin{array}{c}
\mu^0 \\ \mu^1 \\ \mu^2 \\ \mu^3
\end{array}\right]
\label{intensity_sfit_kx}
\end{eqnarray}
and
\begin{eqnarray}
\lefteqn{\cilc\left(\mu,\bmuf\right)=}\nonumber\\
&&\left[\begin{array}{c}
10^{-1}\left(\bmuf\right)^0 \\ 10^{-3}\left(\bmuf\right)^1 \\ 10^{-5}\left(\bmuf\right)^2 \\ 10^{-8}\left(\bmuf\right)^3
\end{array}\right]^T
\left[\begin{array}{rrrr}
-1.08 & 2.09 & -1.78 & 0.66 \\ -0.08 & 6.84 & -11.22 & 6.34 \\ 0.07 & -1.50 & 2.52 & -1.48 \\ -0.06 & 1.03 & -1.90 & 1.18
\end{array}\right]
\left[\begin{array}{c}
\mu^0 \\ \mu^1 \\ \mu^2 \\ \mu^3
\end{array}\right]
\label{linecore_sfit_kx}
\end{eqnarray}
respectively. Since contrast is wavelength dependent and bright magnetic features are largely unresolved at HMI's spatial resolution, these relationships are valid only at the instrument's operating wavelength (6173 \AA) and spatial resolution (1 arcsec).

\begin{figure}[h]
\resizebox{\hsize}{!}{\includegraphics{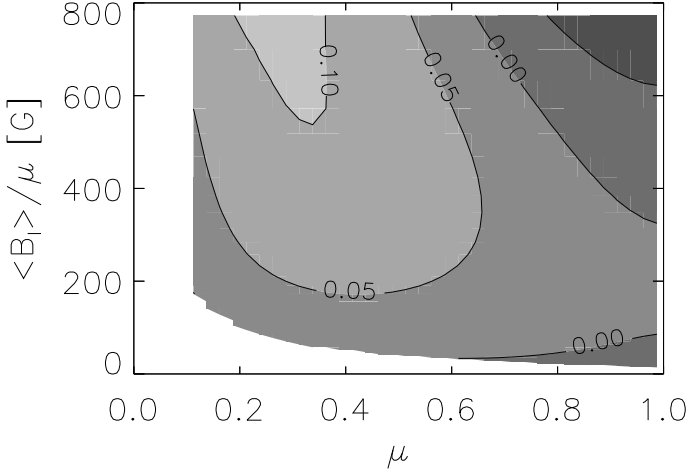}}
\caption{Grey scale plot of the polynomial fit to bin-averaged continuum intensity contrast over the region where data exist, sampled at the mid-point of each bin in the $36\times41$ bins grid employed to compute the averages.}
\label{intensity_contrast_sfit}
\end{figure}

\begin{figure}[h]
\resizebox{\hsize}{!}{\includegraphics{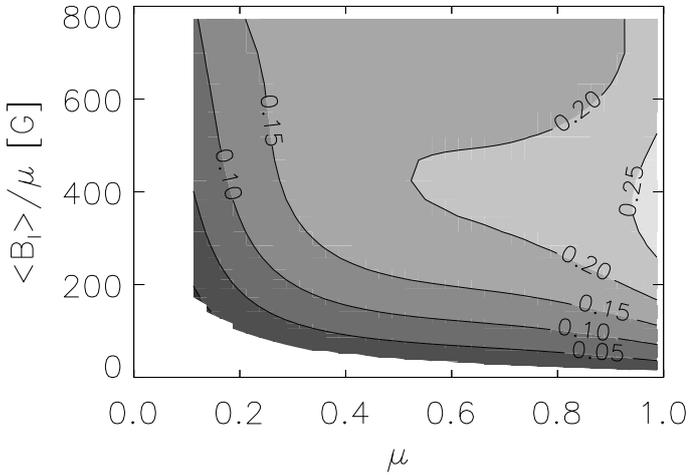}}
\caption{Same as Fig. \ref{intensity_contrast_sfit}, but for line core intensity contrast.}
\label{linecore_contrast_sfit}
\end{figure}

The surface fits are illustrated as surface and grey scale plots in Figs. \ref{intensity_contrast_sfit} and \ref{linecore_contrast_sfit}. Cross-sections to the surface fits are plotted in Figs. \ref{intensity_contrast_scatter_bmu} to \ref{linecore_contrast_scatter_mu} (blue curves) along the contrast CLV and contrast versus $\bmu$ profiles, and the corresponding cubic polynomial fits (red curves). The surface fits are in excellent agreement with the cubic polynomial fits almost everywhere. The agreement is so close that the surface fit cross-sections are completely hidden by the cubic polynomial fits in most places. The advantage with these surface fits is that they allow us to describe how the measured contrasts vary with $\mu$ and $\bmu$ almost equally well and with far fewer free parameters than by the cubic polynomial fit to each individual contrast CLV and contrast versus $\bmu$ profile (32 versus 128 free parameters; 4 from each of 32 profiles).

\subsection{Intrinsic contrast}
\label{intrinsic_contrast}

\begin{figure}[h]
\resizebox{\hsize}{!}{\includegraphics{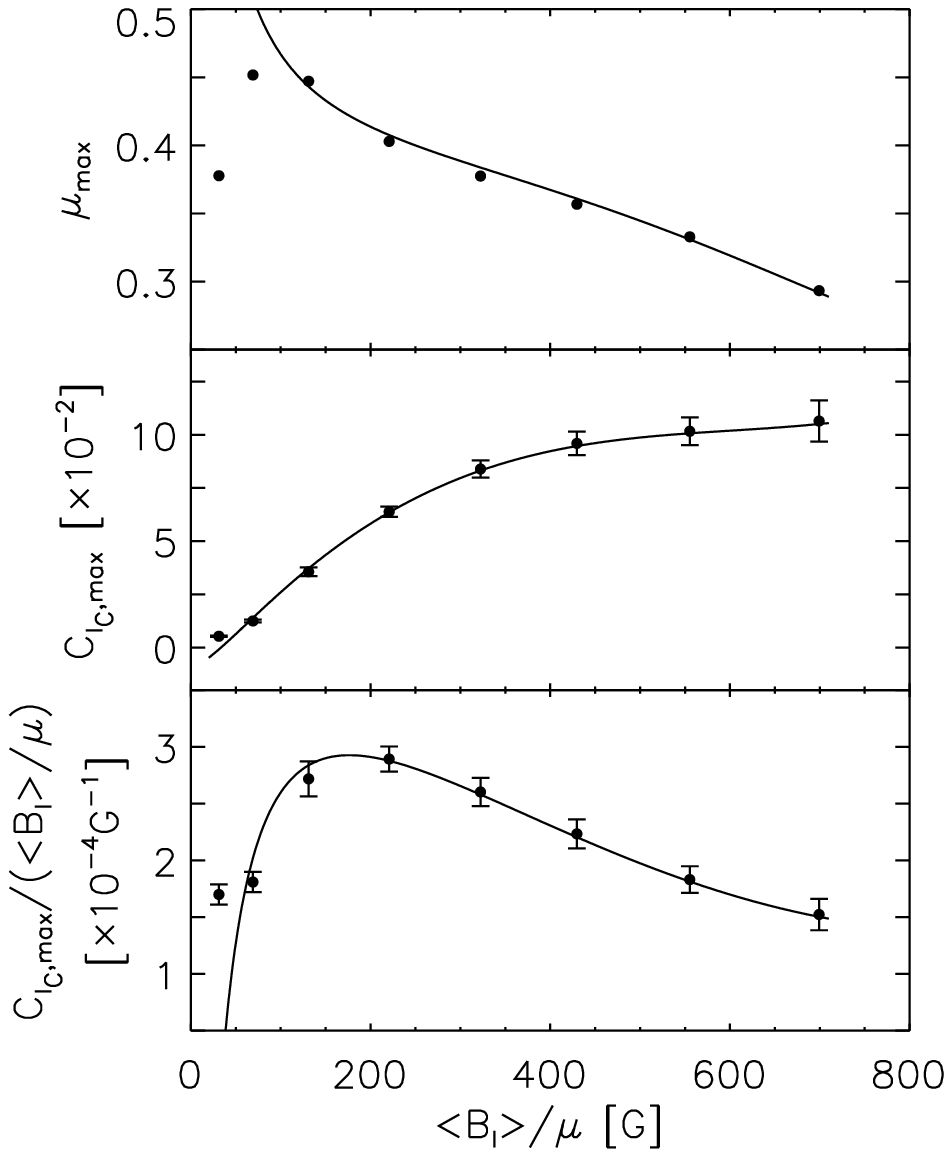}}
\caption{Heliocentric angle at which continuum intensity contrast reaches its maximum, $\umax$ (top), as well as the contrast, $\cicnmax$ (middle) and specific contrast, $\scicnmax$ (bottom) there, as a function of $\bmu$. The filled circles represent the values derived from the cubic polynomial fit to the contrast CLV profiles (Fig. \ref{intensity_contrast_scatter_bmu}) and the error bars the uncertainty in $\cicnmax$ and $\scicnmax$. The curves follow the solution from the surface fit to measured continuum intensity contrast (Eq. \ref{intensity_sfit_kx}).}
\label{intensity_contrast_analysis}
\end{figure}

At each $\bmu$ interval, we estimated the maximum continuum intensity contrast, $\cicnmax$ and the heliocentric angle at which it is reached, $\umax$ from the cubic polynomial fit to the contrast CLV profile (Fig. \ref{intensity_contrast_scatter_bmu}). Following \citet{ortiz02} we term $\scicnmax$ the specific contrast; estimated here from the quotient of $\cicnmax$ and $\mbmu$, the mean $\bmu$ of all the points within a given $\bmu$ interval. The values of $\umax$, $\cicnmax$ and $\scicnmax$ are shown in Fig. \ref{intensity_contrast_analysis} as a function of $\bmu$ where the abscissa is given by $\mbmu$. Also plotted are the values obtained from the surface fit (Equation \ref{intensity_sfit_kx}). The uncertainty in $\cicnmax$ is given by the RMS difference between the contrast CLV profiles and their cubic polynomial fits. The uncertainty in $\scicnmax$ was estimated from the uncertainty in $\cicnmax$ and the standard error of $\mbmu$ employing standard propagation of errors.

\begin{figure}[h]
\resizebox{\hsize}{!}{\includegraphics{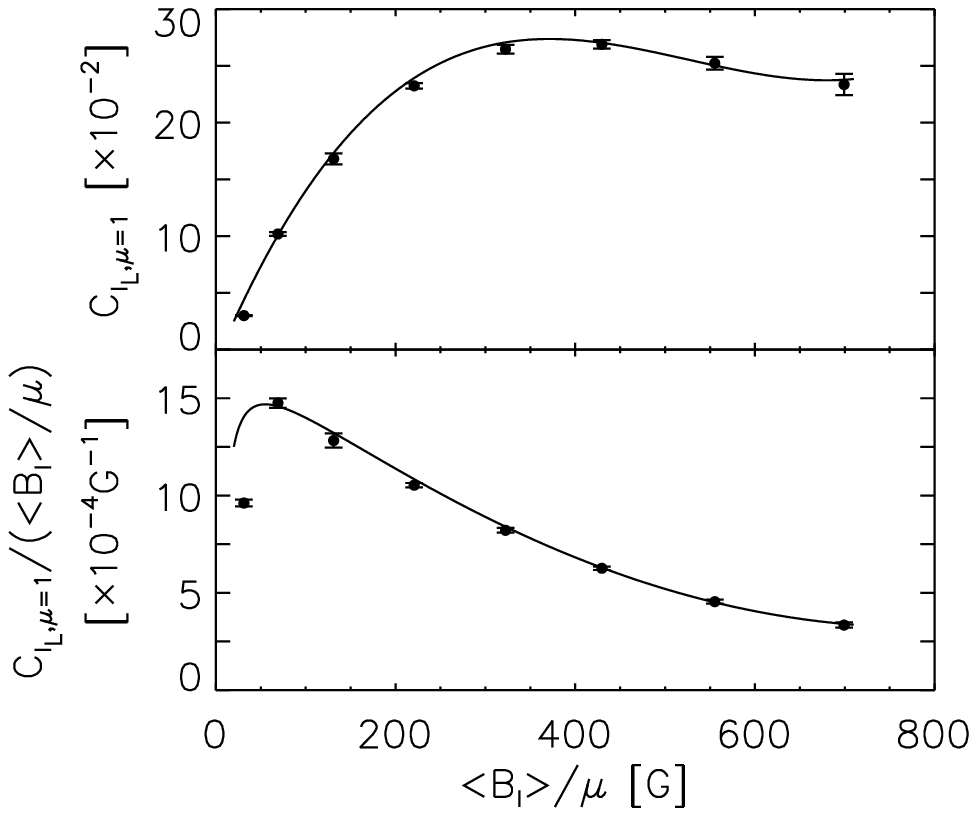}}
\caption{Line core intensity contrast, $\cilcmax$ (top) and specific contrast, $\scilcmax$ (bottom) at disc centre as a function of $\bmu$. The filled circles represent the values derived from the cubic polynomial fit to the contrast CLV profiles (Fig. \ref{linecore_contrast_scatter_bmu}) and the error bars the uncertainty in $\cilcmax$ and $\scilcmax$. The curves follow the solution from the surface fit to measured line core intensity contrast (Eq. \ref{linecore_sfit_kx}).}
\label{linecore_contrast_analysis}
\end{figure}

While the position of the continuum intensity contrast CLV maximum varies with $\bmu$, line core intensity contrast at a given $\bmu$ is invariably strongest at disc centre (Fig. \ref{linecore_contrast_scatter_bmu}) as pointed out in Sect. \ref{profile}. Maximum line core intensity contrast (i.e., the value at $\mu=1$), $\cilcmax$ and specific contrast, $\cilcmax/\left(\bmu\right)$ derived similarly as above from the cubic polynomial fit to the contrast CLV profiles are expressed in Fig. \ref{linecore_contrast_analysis} as a function of $\bmu$, along with the values obtained from the surface fit (Eq. \ref{linecore_sfit_kx}). In both instances, there is some disparity between the values obtained from the cubic polynomial fits to the contrast CLV profiles and the surface fit below $\bmu\sim100\g$. This is likely due to the truncated coverage of disc positions at low $\bmu$ discussed in Sect. \ref{profile}.

Network and facular features exhibit similar, kilogauss strength magnetic fields \citep{stenflo73,solanki84,rabin92,ruedi92}. This means, in general, the intrinsic magnetic field strength of magnetic elements, $B$ does not vary significantly with $\bmu$ and is also well above the range of $\bmu$ examined here, thus allowing us to assume that the magnetic filling factor, $\alpha$ never saturates. As mentioned in Sect. \ref{data}, the quantity $\bmu$ is representative of $\alpha{}B$. Taking these into account, within the limits of this study, $\bmu$ is representative of the magnetic filling factor. Therefore specific contrast, as it is defined and derived here, is a measure of the average intrinsic contrast or heating efficiency of magnetic features within a given $\bmu$ interval. Like the intrinsic field strength, the intrinsic contrast cannot be measured directly because most network and facular features are not resolved at HMI's spatial resolution and due to the effect of scattered light.

Above $\bmu\sim100\g$, the heliocentric angle at which continuum intensity contrast reaches its maximum decreases (i.e., the maximum is reached further away from disc centre) with increasing $\bmu$ (Fig. \ref{intensity_contrast_analysis}). A similar trend was reported by \citet{ortiz02} based on MDI continuum observations. Assuming the hot wall model \citep{spruit76} and a simple flux tube geometry, the authors demonstrated this to imply the average size of flux tubes increases with $\bmu$.

Both continuum and line core intensity specific contrast increases strongly with decreasing $\bmu$ up to a maximum (at around 200 and $50\g$ respectively) before gradually declining (Figs. \ref{intensity_contrast_analysis} and \ref{linecore_contrast_analysis}). The decline at low $\bmu$ aside, the observed trends suggests that brightness or temperature excess in both the lower and middle photosphere in flux tubes decreases with increasing magnetic filling factor \citep[the continuum and line core formation height of the Fe I 6173 \AA{} line is about 20 and 300 km respectively,][]{norton06}. This may partly be due to the increasing average size of flux tubes; larger magnetic elements appear darker as lateral heating is less efficient \citep{spruit76,spruit81,grossmann94} and magnetic suppression of surrounding convective energy transport is more severe \citep{kobel12}. The total radiation flux derived from MHD simulations by \citet{vogler05} exhibits a similar behaviour. This result also agrees with the observations that network elements appear hotter than facular elements in the lower and middle photosphere \citep{solanki84,solanki86,keller90} and G-band bright points are brighter in the quiet Sun than in active regions \citep{romano12}.

The decline in both continuum and line core specific contrast at low $\bmu$ suggests a diminishing heating efficiency of the smallest magnetic elements but is more likely due to the influence of intergranular lanes on apparent contrast. (As with the low contrast towards $\bmuz$ in the contrast versus $\bmu$ profiles discussed in Sect. \ref{profile}.) The probable impact of the truncated coverage of disc positions at low $\bmu$ on the quality of the fits to measured contrast here might have also played a role.

\section{Discussion}
\label{discussion}

\subsection{Comparison with \citet{ortiz02}}
\label{compare_ao}

A comparison of the results reported here for continuum intensity with those from the similar study based on full-disc MDI observations by \citet{ortiz02} reveal several notable differences.
\begin{itemize}
	\item The contrast reported here is generally higher, by as much as a factor of about two. The difference between the two studies becomes increasingly pronounced with $\bmu$ and distance from disc centre.
	\item Approaching disc centre, contrast appears to level off in our study (Fig. \ref{intensity_contrast_scatter_bmu}). In the earlier work, the contrast declines approximately linearly towards $\mu=1$ and there are also marked fluctuations about $\mu\sim0.95$ \citep[Fig. 3,][]{ortiz02}.
	\item Near $\bmuz$ contrast is negative here (Fig. \ref{intensity_contrast_scatter_mu}) but positive in the previous study \citep[Fig. 4,][]{ortiz02}.
	\item The specific contrast (a proxy of intrinsic contrast given by the quotient of contrast at the maximum point on the CLV profile and $\bmu$) presented here ascends with $\bmu$ up to $\bmu\sim200\g$ before descending monotonically thereafter (Fig. \ref{intensity_contrast_analysis}). \citet{ortiz02} found specific contrast to decline approximately linearly with $\bmu$ (Fig. 8 in their paper).
	\item For $\bmu\gtrsim200\g$, the specific contrast reported here is also nearly double that in the earlier work.
\end{itemize}
The lower contrast and specific contrast (for $\bmu\gtrsim200\g$), and difference in CLV towards disc centre reported by \citet{ortiz02} is likely, as we will show shortly, to be primarily due to the misidentification of the magnetic signal adjacent to sunspots and pores as network and faculae (discussed in Sect. \ref{reduction}) by those authors. Care was taken here to minimise such misidentification. The negative contrast towards $\bmuz$ found here, as argued in Sect. \ref{profile}, arises from the resolution of intergranular lanes at HMI's spatial resolution. We demonstrate below that this difference in spatial resolution also contributes to the opposite $\bmu$-dependence of specific contrast below $\bmu\sim200\g$ reported here and by \citet{ortiz02}.

\begin{figure*}[h]
\centering
\includegraphics[width=17cm]{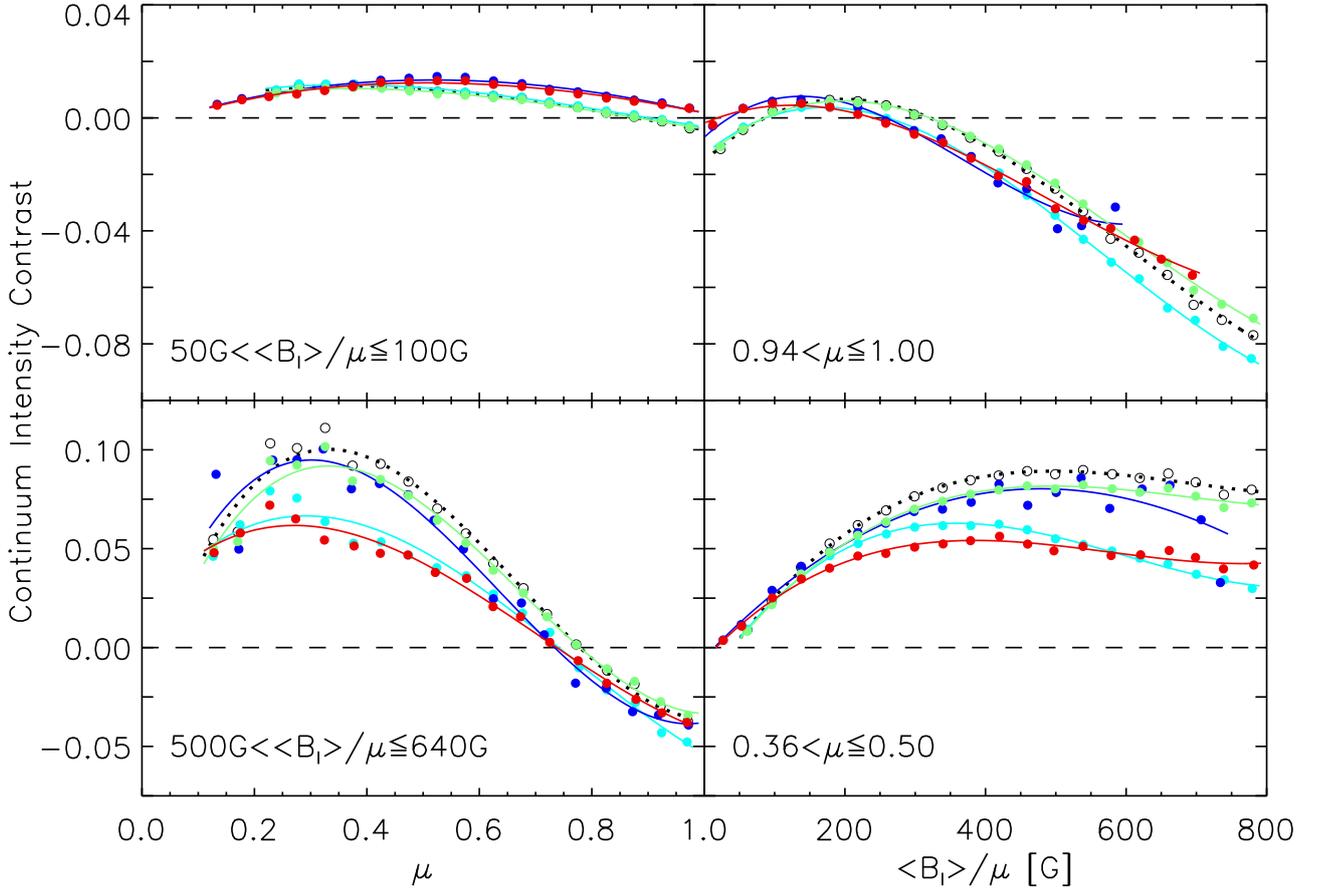}
\caption{Selected continuum intensity contrast CLV (left) and contrast versus $\bmu$ (right) profiles from Figs. \ref{intensity_contrast_scatter_bmu} and \ref{intensity_contrast_scatter_mu} (open circles and dotted curves). The selected profiles correspond to quiet Sun network (top left), active region faculae (bottom left), disc centre (top right) and near limb (bottom right). The cyan, blue and green series denote the profiles obtained by omitting the magnetic extension removal procedure, spatially binning the data set by $4\times4$ pixels and converting measured contrast to 6768 Å respectively. The red series indicates the results of taking into account all these three considerations. The circles represent the mean of measured contrast binned as in the referenced figures and the curves the corresponding third-order polynomial fits. The dashed lines mark the mean quiet Sun level.}
\label{intensity_contrast_canopy}
\end{figure*}

\begin{figure}[h]
\resizebox{\hsize}{!}{\includegraphics{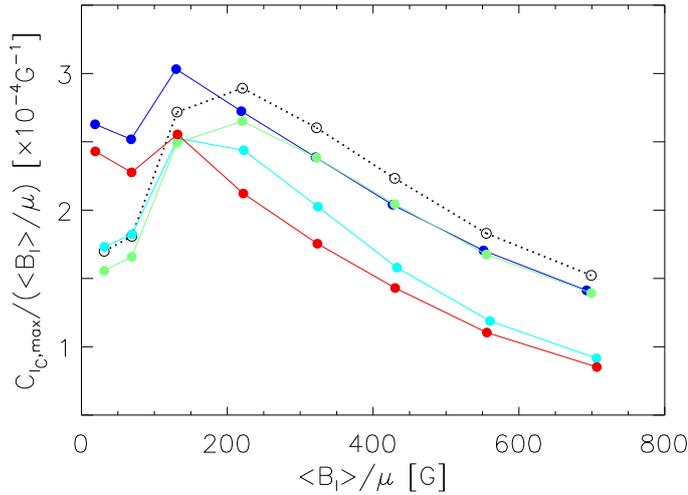}}
\caption{Continuum intensity specific contrast as a function of $\bmu$ from Fig. \ref{intensity_contrast_analysis} (open circles). The cyan, blue and green filled circles represent the same quantity obtained by omitting the magnetic extension removal procedure, spatially binning the data set by $4\times4$ pixels and converting measured contrast to 6768 Å respectively. The red filled circles represent the result of taking into account all these three considerations. The points in each series are joined by straight lines to aid the eye.}
\label{intensity_contrast_analysis_bin}
\end{figure}

To recreate the conditions of the study by \citet{ortiz02}, we recomputed continuum intensity contrast and specific contrast from the HMI data set employed here without applying the magnetic extension removal procedure, binning the data set spatially by $4\times4$ pixels to be consistent with MDI's spatial resolution and (very approximately) transforming measured contrast to the corresponding value at MDI's operating wavelength, 6768 \AA{}. This last transformation was carried out by describing quiet Sun, network and faculae as black bodies, and taking an effective temperature of 5800 K for the quiet Sun. This allowed us to crudely convert contrast measured at 6173 \AA{} by HMI into the corresponding contrast at 6768 \AA{}. It should be noted that this is a first-order approximation of the wavelength dependence of contrast which ignores the variation of the continuum formation height with wavelength \citep{solanki98a,sutterlin99,norton06}. In Fig. \ref{intensity_contrast_canopy} we depict the contrast CLV profile of network patterns ($\bmub$) and active region faculae ($\bmug$) after the application of the above procedure. Also plotted are the similarly treated contrast versus $\bmu$ profiles about disc centre ($\mua$) and near limb ($\mug$). The specific contrasts from this process are illustrated in Fig. \ref{intensity_contrast_analysis_bin}.

Excluding the procedure to remove magnetic signal adjoined to sunspots and pores produced the largest effect. It produced an overall drop in contrast (cyan series, Fig. \ref{intensity_contrast_canopy}) and also reproduced the linear CLV near disc centre observed by \citet{ortiz02} ($\bmuh$ panel). The effects of the resizing of the data set (blue series) and the projection of measured contrasts to MDI's operating wavelength (green series) are relatively minor. The contrast profiles derived from the application of all three procedures (red series) are, in terms of both form and magnitude, in general agreement with the profiles covering similar disc positions and magnetogram signal levels reported by \citet{ortiz02} (Figs. 3 and 4 in their paper).

Similarly for specific contrast, the result of applying all three processes (red series, Fig. \ref{intensity_contrast_analysis_bin}) resembles the measurements presented by \citet{ortiz02} (Fig. 8 in their paper) and the greatest effect on magnitude came from the omission of the magnetic extension removal procedure. Binning the data set down to a MDI-like resolution also produced a significant increase in specific contrast at low $\bmu$. After combining with the other two processes, this rendered specific contrast approximately level below $200\g$. The influence of granulation on apparent contrast at HMI's spatial resolution plays a role in the observed decline in specific contrast from $\bmu\sim200\g$ towards $\bmuz$.

The largest effect is produced by the removal of magnetic signal adjoining to sunspots and pores. Hence it is worth considering this process more closely. The procedure by which we removed these signals inevitably discriminates against active region faculae, and various authors have noted lower contrast in active region faculae compared to quiet Sun network at similar magnetogram signal levels \citep{lawrence93,kobel11}, but we do not reckon this to be the reason for the greater contrast reported here. In spite of the severe steps taken to exclude magnetic signals adjoined to sunspots and pores, there remains a fair representation of active region faculae in the measured contrasts (Fig. \ref{network_plage_mask}). Also, at HMI's spatial resolution, network patterns are largely confined to the lower half of the $\bmu$ range considered (Fig. \ref{qs_ar_fac_map}) while the difference between the contrast reported here and by \citet{ortiz02} is more pronounced at higher $\bmu$.

Following the example of earlier studies \citep[e.g.][]{chapman80}, \citet{ortiz02} fit quadratic polynomials to continuum intensity contrast CLV profiles derived in their study. Here we fit cubic polynomials instead because of the different CLV near disc centre which we just demonstrated to arise from our treatment of magnetic signal adjoined to sunspots and pores.

We surmise that the fluctuations about disc centre in the continuum intensity CLV profiles reported by \citet{ortiz02} is due to the non-homogeneous distribution of active regions over the solar disc. Each unit $\mu$ represents a greater radial distance towards the disc centre rendering these fluctuations increasingly abrupt. Not accounting for magnetic signal adjoined to sunspots and pores probably accentuated these fluctuations.

\subsection{Continuum intensity contrast about disc centre}
\label{compare_dc}

In this section we compare the magnetogram signal dependence of continuum intensity contrast near disc centre reported here with that from other recent studies, summarized in Table \ref{disc_centre_compare}. Apart from the work of \citet{ortiz02}, the CLV of continuum intensity contrast has been examined in detail previously by \citet{topka92,topka97} and \citet{lawrence93}. While we and \citet{ortiz02} employed full-disc observations made at a single wavelength and spatial resolution, these authors collated telescope scans made at multiple wavelengths and resolutions. The magnetogram signal intervals represented by the contrast CLV profiles presented in these papers also differ considerably from those presented here and by \citet{ortiz02}. For these reasons it is not straightforward to make any quantitative comparisons with the contrast CLV reported by these studies, therefore the focus on results for near disc centre here.

\begin{table*}[h]
\caption{Continuum intensity contrast dependence on magnetogram signal near disc centre obtained in the present and from earlier studies.}
\label{disc_centre_compare}
\centering
\begin{tabular}{lccccl}
\hline\hline
Reference & Instrument & Target\tablefootmark{b} & Resolution [Arcsec] & Wavelength [\AA{}] & Results\\
\hline
Present & SDO/HMI & $\mu>0.94$\tablefootmark{c} & $1$ & 6173 & Negative near $\bmuz$. Peak at $\bmu\sim200\g$.\\
1, 2, 3, 4 & SVST & AR & $\gtrsim0.3$ & Various\tablefootmark{d} & Negative near $\bz$. Monotonic decline with $\bl$.\\
3 & SVST & QS & $\gtrsim0.3$ & 6302 & Negative near $\bz$. Peak at $\bl\sim400\g$.\\
 & SFO/SHG\tablefootmark{a} & QS & $\gtrsim1$ & 6302 & Negative near $\bz$. Peak at $100\g\lesssim{}\bl\lesssim200\g$.\\
 & & AR & & & Negative near $\bz$. Monotonic decline with $\bl$.\\
5 & SoHO/MDI & $\mu>0.92$\tablefootmark{c} & $4$ & 6768 & Positive near $\bmuz$. Peak at $\bmu\sim100\g$.\\
6 & SST & AR & $0.15$ & 6302 & Negative near $\bz$. Peak at $\bl\sim650\g$.\\
7 & Hinode/SOT & QS & $0.3$ & 6302 & Negative near $\bz$. Peak at $\bl\sim700\g$.\\
 & & AR & & & Negative near $\bz$. Peak at $\bl\sim700\g$.\\
8 & SST & QS & $0.15$ & 6302 & Negative near $\bz$. Saturate at $\bl\sim800\g$.\\
 & Hinode/SOT & QS & $0.3$ & 6302 & Negative near $\bz$. Peak at $\bl\sim500\g$.\\
\hline
\end{tabular}
\tablebib{(1) \citet{title92}; (2) \citet{topka92}; (3) \citet{lawrence93}; (4) \citet{topka97}; (5) \citet{ortiz02}; (6) \citet{narayan10}; (7) \citet{kobel11}; (8) \citet{schnerr11}.}
\tablefoot{
\tablefoottext{a}{San Fernando Observatory (SFO) 28-cm vacuum telescope and vacuum spectroheliograph (SHG).}
\tablefoottext{b}{AR and QS denote active region and quiet Sun respectively.}
\tablefoottext{c}{Segment of full-disc observations considered.}
\tablefoottext{d}{5250\AA{}, 5576\AA{}, 6302\AA{} and 6768\AA{}.}
}
\end{table*}

The negative contrast at low magnetogram signal levels found here and by the majority of the compared works counters expectations from thin flux tube models, which predict intrinsically bright magnetic features in this regime \citep{knoelker88}. As discussed in Sect. \ref{profile}, this is attributed here \citep[and in][]{title92,topka92,kobel11} to the influence of intergranular lanes, an assertion supported by various models \citep{title96,schnerr11}. Of the studies compared, only that by \citet{ortiz02} noted positive contrasts. Binning the data set here to a MDI-like resolution raised the disc centre end of the contrast CLV profile in the $\bmub$ interval into positive territory (Fig. \ref{intensity_contrast_canopy}), suggesting that the positive contrasts reported by \citet{ortiz02} arose from utilising data at a resolution (the lowest of the studies compared) where granulation is largely unresolved.

From SVST scans of active regions, \citet{title92}, \citet{topka92,topka97} and \citet{lawrence93} found contrast to decline monotonically with magnetogram signal. \citet{lawrence93} also noted the same with active region data acquired at the San Fernando Observatory (SFO). \citet{kobel11} demonstrated the comparatively poorer resolution and straylight from pores to be likely culpable for the monotonic decline observed in the SVST studies. In all the other compared works, contrast exhibits a peak, the position of which varied from $\bmu\sim100\g$ \citep{ortiz02} to $\bl\sim700\g$ \citep{kobel11}, except for the quiet Sun SST scan examined by \citet{schnerr11}, where contrast saturated at $\bl\sim800\g$. Generally, the finer the spatial resolution, the higher the position of the peak. Most of these studies were based on observations made at the Fe I 6302 \AA{} line. Even if we discount the studies made at other wavelengths, this broad pattern is still apparent, ruling out differences in wavelength as the major driver. \citet{kobel11} reported similar contrast peak positions for quiet Sun and active region scans. Therefore the difference between active region and quiet Sun contrast had likely little role in the spread in reported peak positions amongst the compared works. As shown in the $\mua$ panel of Fig. \ref{intensity_contrast_canopy}, both resizing the HMI data set to a MDI-like spatial resolution and omitting the magnetic extension removal procedure shifted the position of the contrast versus $\bmu$ profile maximum towards the origin.

The above comparison points to differences in spatial resolution and treatment of magnetic signal near sunspots and pores as the dominant factors behind the spread in reported dependence of contrast on magnetogram signal. Indeed, the recent MHD simulation work of \citet{rohrbein11} suggests that the contrast peak at intermediate magnetogram signal levels seen in direct observations, but not observed in MHD simulations, is a product of the limited spatial resolution. The bulk of the studies listed in Table \ref{disc_centre_compare} were based on higher spatial resolution data than utilised here. For this reason, a more quantitative comparison like we did in Sect. \ref{compare_ao} with the findings of \citet{ortiz02} is not workable here.

The recent works of \citet{berger07} and \citet{viticchie10}, examining the contrast of bright points in the G-band, bear tenuous relevance to our study. These studies utilised observations made in a molecular band (i.e., neither continuum nor line core) where the contrast of magnetic features is enhanced \citep[due to CH depletion,][]{steiner01,schussler03}. More importantly, while this study and the works cited in Table \ref{disc_centre_compare} considered each pixel a separate entity, \citet{berger07} and \citet{viticchie10} examined the overall contrast of each individual bright structure. As the body of magnetic features isolated by both approaches differ, the results are not directly comparable. \citet{berger07} did however also report a pixel-by-pixel consideration of G-band contrast. Scans at four disc positions were surveyed. Barring one that appeared anomalous, the contrast versus magnetogram signal profiles from each scan bear general resemblance to ours in terms of form. Notably, the profile from the disc centre scan, with a spatial resolution of $0.15\:{\rm arcsec}$, exhibits a peak at $\bl\sim700\g$, consistent with the broad pattern between spatial resolution and peak position described here.

\subsection{Line core intensity contrast}
\label{compare_lc}

As pointed out in Sect. \ref{profile}, the most notable difference between the continuum and line core intensity contrasts present here is the converse CLV. While continuum intensity contrast is weakest near disc centre and strengthens towards the limb, line core intensity contrast is strongest at disc centre and declines towards the limb. The divergent CLV exhibited by the two sets of measurements stem from their rather different physical sources. Continuum intensity is enhanced largely in the hot walls of magnetic elements, thus the centre-to-limb increase. The line core is formed in the middle photosphere, which is heated either by radiation from deeper layers \citep{knoelker91}, or by mechanical and Ohmic dissipations \citep{moll12}. Also stated in Sect. \ref{profile}, line core intensity is modulated by line strength and shape, and continuum intensity. Excluding variation related to continuum excess, line core intensity enhancement arises from the influence of the temperature excess in the middle photosphere and Zeeman splitting on line strength and shape. Given the relatively narrow, vertical geometry of flux tubes, as magnetic elements rotate from disc centre to limb, line-of-sight rays go from being largely confined to single flux tubes to increasingly passing into and out of multiple flux tubes, especially in densely packed facular regions \citep{bunte93}. Line core intensity contrast decreases towards the limb as the contribution to the spectral line from magnetic elements diminishes from absorption in the non-magnetic part of the atmosphere transversed by the rays \citep{solanki98b}. Another probable cause of the centre-to-limb decline is the spatial displacement of the line core with respect to the corresponding continuum towards the limb caused by the difference in formation height and oblique viewing geometry \citep{stellmacher91,stellmacher01}. The line core intensity enhancement arising from temperature excess in the middle photosphere and Zeeman splitting discussed here is not to be confused with that from the centre-to-limb broadening of the Fe I 6173 \AA{} line mentioned in Sect. \ref{qsintensity}, which arises from the viewing geometry independent of magnetic field.

\begin{figure*}[h]
\centering
\includegraphics[width=17cm]{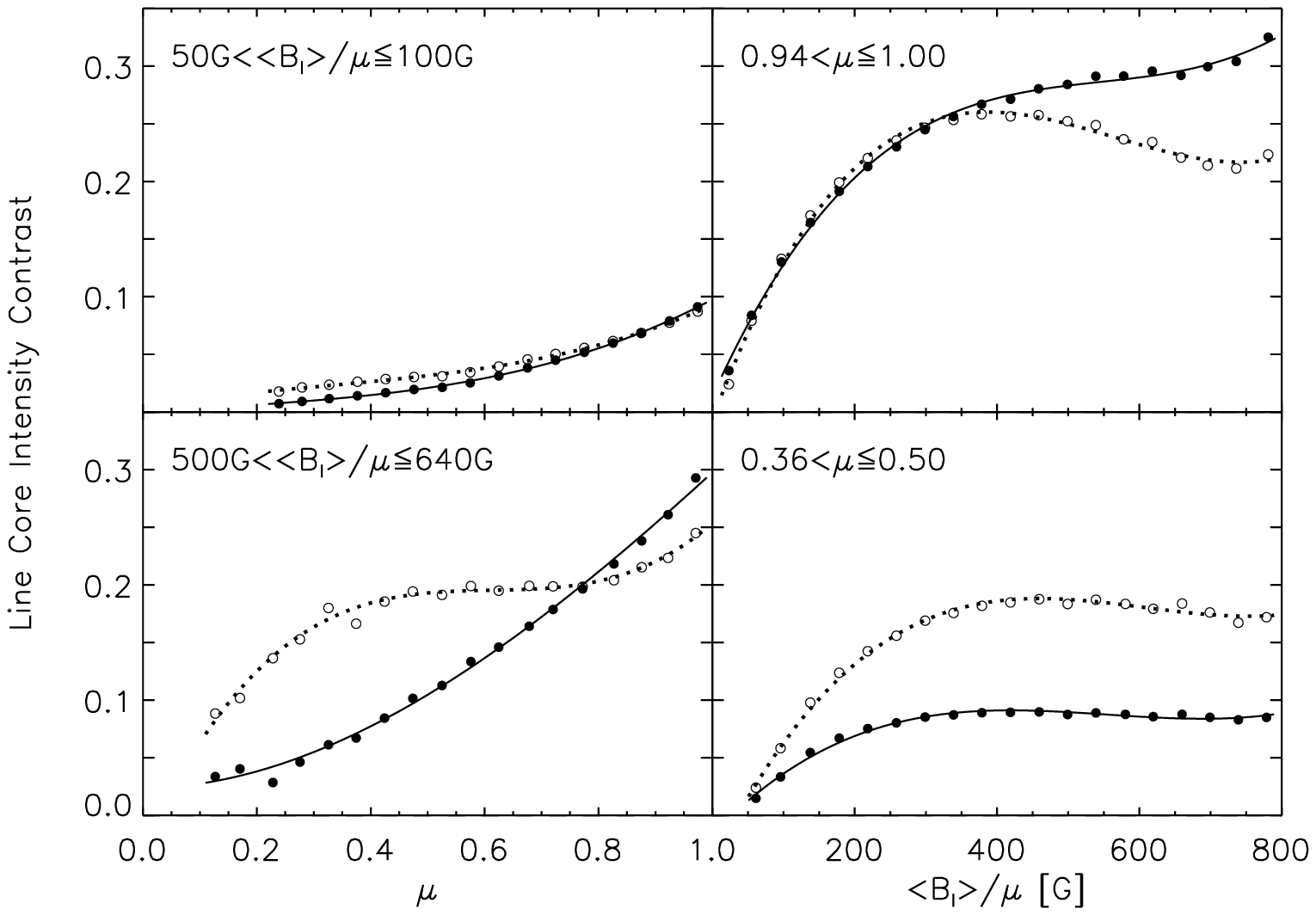}
\caption{Selected line core intensity (open circles and dotted curves) and corresponding residual intensity (filled circles and solid curves) contrast CLV (left) and contrast versus $\bmu$ (right) profiles. The circles represent the mean of measured contrast binned as in Figs. \ref{linecore_contrast_scatter_bmu} and \ref{linecore_contrast_scatter_mu} and the curves the corresponding cubic polynomial fits. The selected profiles correspond to quiet Sun network (top left), active region faculae (bottom left), disc centre (top right) and near limb (bottom right).}
\label{line_weakening}
\end{figure*}

\begin{figure}[h]
\resizebox{\hsize}{!}{\includegraphics{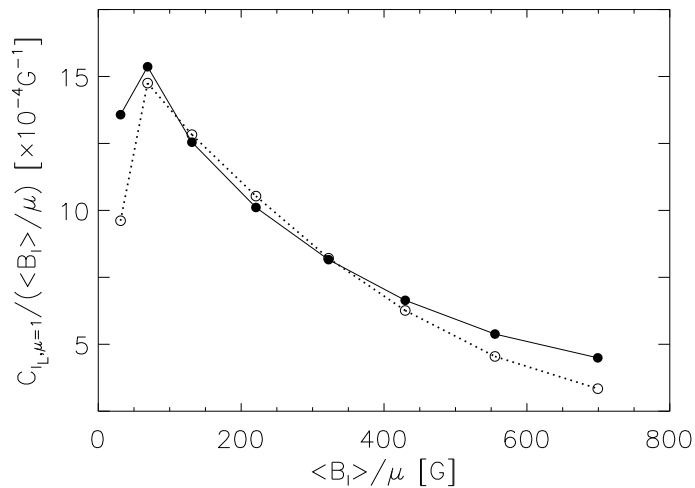}}
\caption{Line core intensity (open circles) and corresponding residual intensity (filled circles) specific contrast as a function of $\bmu$. The circles are joined by straight lines to aid the eye.}
\label{linecore_contrast_analysis_bin}
\end{figure}

We recomputed line core intensity contrast, this time normalizing the line core intensity images by the corresponding continuum intensity images prior to data reduction. The result, the line core residual intensity contrast, is essentially the component of line core intensity contrast arising from line weakening in magnetic features alone. Line core intensity and residual intensity contrast values can be compared directly. (Line core residual intensity contrast equates to the line core intensity contrast we would get from scaling the line core intensity of just the network and faculae pixels by $\mqsicn/\icn$.) The line core residual intensity contrast CLV profile of network patterns ($\bmub$) and active region faculae ($\bmug$), and contrast versus $\bmu$ profile about disc centre ($\mua$) and near limb ($\mug$) so derived are plotted along with the corresponding line core intensity contrast profiles from Figs. \ref{linecore_contrast_scatter_bmu} and \ref{linecore_contrast_scatter_mu} in Fig. \ref{line_weakening}. Line core residual intensity specific contrast is plotted along with the line core intensity specific contrast from Fig. \ref{linecore_contrast_analysis} in Fig. \ref{linecore_contrast_analysis_bin}.

As expected, excluding the contribution of continuum intensity enhancement to line core intensity contrast results in a more pronounced centre-to-limb decline ($\bmub$ and $\bmug$ panels, Fig. \ref{line_weakening}). This decline in the line core residual intensity contrast CLV profiles is also consistent with the spectroscopic observations of \citet{stellmacher79} and \citet{hirzberger05}. Comparing line core intensity and residual intensity contrast (Figs. \ref{line_weakening} and \ref{linecore_contrast_analysis_bin}), the broad similarity, especially near disc centre, imply that line core intensity contrast is dominated by the contribution from line weakening. Even towards the limb, where magnetic line weakening is at its weakest, it still comprises a significant proportion of observed line core intensity contrast. The potential implications of this result and the opposite CLV exhibited by continuum and line core intensity contrast on facular contribution to solar irradiance variations will be discussed in Sect. \ref{fac_tsi}.

Very few studies examining the disc position and magnetogram signal dependence of line core intensity contrast exist in the literature. \citet{frazier71}, \citet{lawrence91} and \citet{title92} measured intensity contrast at disc centre in the core of the Fe I 5250 \AA{}, Fe I 6302 \AA{} and Ni I 6768 \AA{} lines, respectively. Results from different lines are not directly comparable but it is still encouraging to see that in terms of magnitude and general trend with magnetogram signal, the results from these earlier studies express broad agreement with ours. The only notable exception is the steep monotonic decline at $\bl>600\g$ reported by \citet{title92}, which largely persisted even after the authors masked out pores. The decline coincides with a similar trend in intensity contrast measured at the nearby continuum, suggesting a greater relative influence of the continuum excess to magnetic line weakening at the Ni I 6768 \AA{} line compared to the Fe I 6173 \AA{} line. \citet{frazier71} also made measurements away from disc centre, but due to data scatter the author could do no better than to present a schematic representation of the CLV through visually fitted linear functions. \citet{walton87} reported measured line core residual intensities for eight magnetically insensitive lines at various disc positions. There were however only a small number of scattered measurements for each line, and a relatively narrow range of disc positions ($\mu>0.6$). To our knowledge, the present study is the first to examine line core intensity contrast employing full-disc observations. Together with the relatively fine resolution and low noise level, this allowed us to describe the magnetogram signal dependence and especially the CLV with greater accuracy and detail than in the previous efforts.

\subsection{Facular contribution to variation in solar irradiance}
\label{fac_tsi}

Our finding that intensity contrast in the line core is dominated by line weakening rather than continuum intensity enhancement and exhibits the opposite CLV as continuum intensity contrast has potential implications on facular contribution to total solar irradiance variation. Total solar irradiance (TSI) variation is the sum manifest of variation in the continuum and spectral lines. The converse CLV of the continuum and line core intensity contrasts reported here imply a different time variation of the contribution by the continuum and spectral lines to solar irradiance as magnetic features rotate across the solar disc.

\begin{figure*}[h]
\centering
\includegraphics[width=17cm]{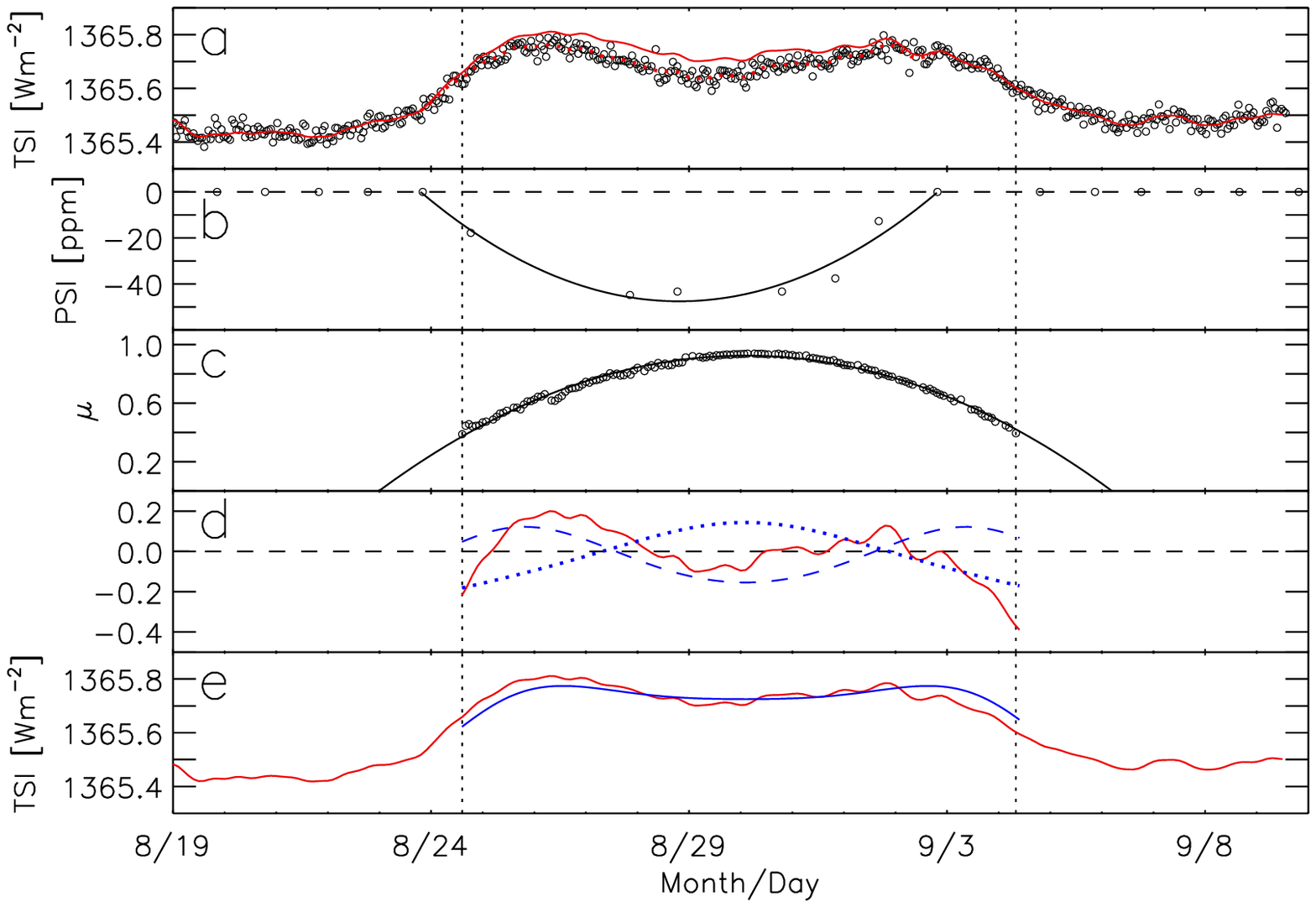}
\caption{a) Total solar irradiance, TSI from DIARAD on SoHO/VIRGO (open circles) for the period of August 19 to September 9, 1996. The dotted and solid red curves represent the interpolated and smoothed version before and after subtracting sunspot darkening. b) Photometric Sunspot Index, PSI (open circles) by \citet{chapman94} and the quadratic polynomial fit to the non-zero points (curve). The dashed line marks the zero level. c) Position of NOAA 7986 as estimated from MDI magnetograms (open circles) and the quadratic polynomial fit (curve). d) TSI minus sunspot darkening (red curve) and model of facular contribution to TSI based on observed intensity contrast in the continuum (blue dashed curve) and line core (blue dotted curve), each mean subtracted and normalized by the area bound between the time series and the zero level (black dashed line). e) TSI minus sunspot darkening (red curve) and the multiple linear regression fit of the continuum and line core models (blue solid curve). The dotted lines running down all the panels mark the period where NOAA 7986 was entirely on the solar disc in MDI magnetograms.}
\label{lone_ar}
\end{figure*}

In Fig. \ref{lone_ar}a we show the level 2.0 hourly TSI measurements from the DIARAD radiometer \citep{dewitte04} on the SoHO/VIRGO instrument \citep{frohlich95} for the 22-day period of August 19 to September 9, 1996 (open circles). Active region NOAA 7986 rotated into view on August 23 and out of view on September 5, and was the only active region on the solar disc for the duration of its passage across the solar disc. The period is otherwise relatively quiet. This data therefore allows us to chart variations in TSI arising mainly from the passage of a single active region across the solar disc. The dip around August 29 corresponds to active region NOAA 7986 crossing disc centre. Even with darkening from sunspots present in the active region, the nadir of the dip is $\sim0.2\:{\rm Wm^{-2}}$ above the level before August 23 and after September 5. This suggests an overall positive contribution to TSI variation by the faculae in NOAA 7986 when it was near disc centre \citep[first pointed out by][]{fligge00}.

A schematic representation of facular contribution to variation in TSI during the passage of NOAA 7986 across the solar disc, depicted in Fig. \ref{lone_ar}e, was derived as follows.
\begin{itemize}
	\item The DIARAD TSI data was interpolated at 0.1 day intervals and the result smoothed \citep[via binomial smoothing,][]{marchand83}. Sunspot darkening was estimated from the Photometric Sunspot Index (PSI) by \citet{chapman94} based on full-disc photometric images acquired with the Cartesian Full Disk Telescope 1 (CFDT1) at SFO. A quadratic polynomial was fitted to the PSI values from the period of interest excluding the points where ${\rm PSI}=0$ (i.e., no sunspots in view). Taking a value of $1365.4\:{\rm Wm^{-2}}$ for the total irradiance of the quiet Sun, the fit was converted from units of parts per million to ${\rm Wm^{-2}}$ and subtracted from the DIARAD data. This value for the total irradiance of the quiet Sun is given by the average TSI at the last three solar minima stated in version d41\_62\_1204 (dated April 2, 2012) of the PMOD TSI composite \citep{frohlich00}. The DIARAD data, after this treatment, represents variation in TSI largely from faculae in NOAA 7986 alone. In Fig. \ref{lone_ar}a we plot the DIARAD data after interpolation and smoothing (red dotted curve), and after removing sunspot darkening (red solid curve) along the original measurements. In Fig. \ref{lone_ar}b we show the PSI (open circles) and the quadratic polynomial fit to the non-zero segment (curve).
	\item The trajectory of NOAA 7986 during its passage across the solar disc was estimated from 142 level 1.8 5-minute MDI magnetograms \citep{scherrer95} on which the active region was entirely in view (i.e. not only partially rotated into, or partially rotated off the solar disc). Taking the unsigned magnetogram signal, the magnetograms were binned spatially by $16\times16$ pixels. For each binned magnetogram, the position of NOAA 7986, in terms of $\mu$, was estimated from the mean position of the five pixels within the active region with the strongest signal. The trajectory of NOAA 7986 is then given by the quadratic polynomial fit to these estimates. In Fig. \ref{lone_ar}c we show the estimated position of NOAA 7986 in the magnetograms (open circles) and the quadratic polynomial fit (curve).
	\item Facular contribution to variation in TSI was very approximately modelled from the empirical relationships describing contrast as a function of $\mu$ and $\bmu$ derived in this study. Assuming a power law distribution of $\bmu$ with a scaling exponent of -1.85 \citep{parnell09}, we evaluated Eqs. \ref{intensity_sfit_kx} and \ref{linecore_sfit_kx}, scaled by $\left(\bmuf/15\right)^{-1.85}$ and integrated over $\bmu=15\g$ to $800\g$, at 0.1 day intervals taking $\mu$ from the trajectory of NOAA 7986 estimated earlier. The resulting time series were then scaled by $\mu$ (to correct for the CLV of projected area on the solar disc) and the limb darkening function from \citet{foukal04}. Given the approximate manner of this derivation, the results are non-quantitative. However for this analysis it is not the actual values but the temporal trends that is important. Here we had approximated the active region as a point object and so cannot include variation arising from the active region being only partially visible as it rotates into and off the solar disc. This derivation is therefore only valid, and confined to, the period where NOAA 7986 was entirely in view in MDI magnetograms.
\end{itemize}
The treated DIARAD data, giving TSI variation largely from faculae in NOAA 7986 alone, is plotted in Fig. \ref{lone_ar}d (red curve) along the conjectures based on the observed intensity contrast in the continuum (blue dashed curve) and line core (blue dotted curve). To compare how they varied with time, we subtracted the mean from and normalized each time series by the area bounded by the curve and the zero level. In Fig. \ref{lone_ar}e we show the multiple linear regression fit of the continuum and line core models to the DIARAD series (blue solid curve), which showed a much better agreement to it than either model. Facular contribution to variation in solar irradiance appears to be strongly driven by intensity contrast in both the continuum, and spectral lines, which derives largely from line weakening in magnetic elements. This complies with the observation that solar irradiance variations over the solar cycle seems to be significantly influenced by changes in spectral lines \citep{mitchell91,unruh99,preminger02}.

Various studies examining the photometric contrast of faculae, as identified in calcium line images, reported positive values near disc centre \citep{lawrence88,walton03,foukal04}. The apparent divergence between the findings of these works, and the largely negative contrast near disc centre reported here (Fig. \ref{intensity_contrast_scatter_mu}) and in the similar studies listed in Table \ref{disc_centre_compare}, where network and faculae were characterized by the magnetogram signal, arises from the different selection methods \citep{ermolli07}. Significantly, the results of \citet{foukal04} were based on near total light broadband measurements. From the statistical analysis of an extensive catalogue of full-disc photometric observations obtained with the SFO/CFDT1, \citet{walton03} found contrast, in a continuum passband (i.e., negligible spectral line contribution) in the visible red, averaged over each facular region (as identified in the Ca II K) to be positive ($\sim0.005$) near disc centre. The authors took this result to indicate the distribution of flux tube sizes in facular regions is biased to the low end (i.e., more small bright flux tubes than large dark ones), which is reasonable considering the typical distribution of magnetogram signal \citep[i.e., much more weak signals than strong,][]{parnell09} and the observation that the average flux tube size is greater where magnetogram signal is greater \citep{ortiz02}. This suggests the continuum component of facular contribution to TSI variation is on average positive at disc centre in the visible red. From this and the analysis presented above, we conclude that the apparent overall positive contribution to TSI variation by the faculae in NOAA 7986 when it was near disc centre is, at least in part, due to the prevalence of smaller flux tubes and line weakening.

\section{Conclusion}
\label{conclusion}

Here we have presented measured network and facular intensity contrast in the continuum and in the core of the Fe I 6173 \AA{} line from SDO/HMI full-disc observations. We studied the dependence of the contrast on disc position and magnetogram signal, represented by $\mu$ and $\bmu$ respectively. Specifically, we derived empirical relationships describing contrast as a function of $\mu$ and $\bmu$, and specific contrast (contrast per unit $\bmu$, representative of intrinsic contrast) as a function of $\bmu$. This study exploits the unprecedented opportunity offered by the SDO mission to examine co-temporal full-disc observations of magnetic field and intensity at a constant intermediate spatial resolution (1 arcsec), relatively low noise, and without atmospheric interference. The quality of the data allowed us to examine intensity contrast for a larger sample at greater accuracy and detail than previous, similar studies, especially in the case of the line core. These results constitute stringent observational constraints on the variation of network and facular intensity with disc position and magnetogram signal in the low and middle photosphere. By constraining atmospheric models of network and faculae, these results should be of utility to solar irradiance reconstructions, especially as HMI data will increasingly be used for this purpose. Given this is the first study of its kind to examine the entire solar disc in both the continuum and line core, it should also be useful to reproduce these results in models of magnetic flux concentrations.

There are significant discrepancies in the continuum intensity contrast reported here and from earlier studies. In this study we had taken steps to account for magnetic signal in the periphery of sunspots and pores, arising from their magnetic canopies and the influence of straylight, which can easily be misidentified as network and faculae. From a comparison with the findings of past efforts, including a recomputation of the results obtained here recreating the conditions of the similar study by \citet{ortiz02}, we showed differences in resolution, and treatment of magnetic signal adjacent to sunspots and pores to be the likely main factors behind the spread in reported results. The apparent radiant behaviour of network and faculae elements is strongly modulated by spatial resolution \citep{title96,rohrbein11,schnerr11}. An understanding of its influence is necessary for the proper interpretation of direct measurements.

In terms of magnitude, trend with magnetogram signal and in particular the CLV, the results obtained here in the continuum and line core differ considerably. While continuum intensity contrast broadly ascends towards the limb, line core intensity contrast is greatest near disc centre and diminishes from disc centre to limb. The divergence between both sets of measurements arises dominantly from spectral line changes due to heating in the middle photosphere and Zeeman splitting in magnetic features, and the different mechanisms by which apparent contrast vary with viewing geometry going from disc centre to limb. From a simple model based on the empirical relationships between contrast, and $\mu$ and $\bmu$ derived here we confirmed that facular contribution to variation in solar irradiance is strongly driven by both continuum excess and spectral line changes.

The specific contrast in both the continuum and line core exhibit a marked decline with increasing magnetogram signal, confirming that network elements are, per unit magnetic flux, hotter and brighter than active region faculae. The different radiant behaviour of network and faculae, not accounted in present models of solar irradiance variations, would be an important factor to consider for more realistic modelling. This observation also implies that secular changes in solar irradiance may be considerably larger than what some models of solar irradiance variations have suggested, given the variation in the number of small-scale magnetic elements on the solar disc is a prime candidate driver of secular changes \citep{solanki02}. For example, the model employed by \citet{krivova07} to reconstruct variation in TSI from 1700 and by \citet{vieira11} for over the Holocene assumed the faculae contrast model by \citet{unruh99} for both network and faculae. This renders network and faculae with similar magnetic filling factors equally bright and therefore possibly underestimate the contribution by network to secular variation.

\begin{acknowledgements}
KLY acknowledges postgraduate fellowship of the International Max Planck Research School on Physical Processes in the Solar System and Beyond. The authors would like to express their gratitude to the SDO/HMI team for providing the data and members of the German Data Center for SDO (GDC-SDO) team, in particular R. Burston, for assistance rendered over the course of this work. This work has been partly supported by WCU grant No. R31-10016 funded by the Korean Ministry of Education, Science and Technology.
\end{acknowledgements}

\end{document}